\documentclass[12pt]{article}

\usepackage{amsmath}            
\usepackage{amssymb}            %
\usepackage{amsfonts}           %
\usepackage{amsthm}             %

\usepackage{graphicx}           
\usepackage[utf8]{inputenc}     
\usepackage{aas_macros}				  
\usepackage{bm}                 
\usepackage[nopatch=footnote]{microtype} 
\usepackage{etoolbox}           
\usepackage[T1]{fontenc}        

\usepackage{slashed}				
\usepackage{mathrsfs}				
\usepackage{bbm}					  
\usepackage{cancel}					
\usepackage[normalem]{ulem} 
\usepackage{youngtab}	    	
\usepackage{mleftright}     

\usepackage[dvipsnames,table]{xcolor}
\usepackage[hang,flushmargin,stable]{footmisc} 

\usepackage{fancyhdr}		
\usepackage{lipsum}			
\usepackage[most]{tcolorbox} 
\usepackage{subcaption}	
\usepackage{cite}			  
\usepackage{wrapfig}    
\usepackage{float}      

\usepackage{booktabs}		
\usepackage{nicefrac}		
\usepackage{multirow}		
\usepackage{arydshln}		

\usepackage[font=small]{caption} 	
\usepackage{float}         			  
\usepackage{ccicons}              

\usepackage[margin=2.5cm]{geometry} 
\usepackage{changepage}             
\numberwithin{equation}{section}    
\usepackage{marginnote}             

\usepackage{multicol}
\usepackage{relsize}
\patchcmd{\thebibliography}
  {\list}
  {\begin{multicols}{2}\smaller\list}
  {}
  {}
\appto{\endthebibliography}{\end{multicols}}

\let\oldenumerate\enumerate
\renewcommand{\enumerate}{
  \oldenumerate
  \setlength{\itemsep}{4pt}
  \setlength{\parskip}{0pt}
  \setlength{\parsep}{0pt}
}

\let\olditemize\itemize
\renewcommand{\itemize}{
  \olditemize
  \setlength{\itemsep}{4pt}
  \setlength{\parskip}{0pt}
  \setlength{\parsep}{0pt}
}

\newcommand{\email}[1]{\href{mailto:#1}{#1}}

\fancypagestyle{firststyle}{
  \rhead{\footnotesize%
  \texttt{\FlipTR}%
  }}

\usepackage{etoc}

\usepackage{scalefnt} 
\newcommand\acro[1]{{\scalefont{.95}{#1}}} 

\renewcommand{\text}{\textnormal}	        

\usepackage{undertilde} 

\newcommand{\inv}{^{-1}}

\newcommand{\D}[2][]{\ensuremath{\operatorname{d}\mkern-3mu^{#1}\mkern-1mu{#2}}}
 
\usepackage{scalerel} 

\newcommand*{\defeq}{\mathrel{\vcenter{\baselineskip0.5ex \lineskiplimit0pt
                     \hbox{\scriptsize.}\hbox{\scriptsize.}}}%
                     =}

\usepackage{pifont}
\newcommand\Vtextvisiblespace[1][.3em]{%
  \mbox{\kern.06em\vrule height.3ex}%
  \vbox{\hrule width#1}%
  \hbox{\vrule height.3ex}}

\newcommand{\eqfig}[1]{%
  \vcenter{\hbox{#1}}}

\usepackage{xparse}
\ExplSyntaxOn
\NewDocumentEnvironment{nolabel}{}{
  \cs_set_eq:NN \label \use_none:n
  \cs_set_eq:cN { ltx@label} \use_none:n
}{}
\ExplSyntaxOff

\usepackage{array}
\newcolumntype{L}{!{\hspace{2em}}l}

\usepackage{lmodern}

\newcommand{\VBF}{\acro{VBF}}

\def\BibTeX{{\rm B\kern-.05em{\sc i\kern-.025em b}\kern-.08em
    T\kern-.1667em\lower.7ex\hbox{E}\kern-.125emX}{}}

\def\BibLaTeX{{\rm B\kern-.05em{\sc i\kern-.025em b}\kern-.08em
    \LaTeX}{}}

\usepackage[scaled]{helvet} 
\usepackage[T1]{fontenc}    
\usepackage{titlesec}       

\titleformat{\section}{\normalfont\sffamily\Large\bfseries}{\thesection}{1em}{}
\titleformat{\subsection}{\normalfont\sffamily\large\bfseries}{\thesubsection}{1em}{}

\titleformat{\paragraph}[runin]
  {\normalfont\sffamily\normalsize\bfseries} 
  {\theparagraph}                            
  {1em}                                      
  {}                                         

\usepackage[
	colorlinks=true,
	citecolor=green!50!black,
	linkcolor=NavyBlue!75!black,
	urlcolor=green!50!black,
	hypertexnames=false]{hyperref}

\usepackage{orcidlink}			

\graphicspath{{figures/}}           

\begin{document}

\newcommand{\MF}[1]{\textcolor{red}{[\textbf{MF}: #1]}}
\newcommand{\FlipTR}{UCR-TR-2025-FLIP-NCC-1031-A} 
\thispagestyle{firststyle}


\begin{center}
    {\huge \textbf{Dark $Z'$ at a Muon Collider}: \par}
    {\Large \textbf{Radiative Return versus Vector Boson Fusion} \par}
    \vskip .5cm
\end{center}




\newcommand{\authorA}{Philip Tanedo}
\newcommand{\emailA}{flip.tanedo@ucr.edu}
\newcommand{\orcidA}{0000-0003-4642-2199}
\newcommand{\institutionA}{
		Department of Physics \& Astronomy, 
	    University of  California, Riverside, 
	    CA 92521, USA}

\newcommand{\authorB}{Marvin Flores}
\newcommand{\emailB}{mflores@nip.upd.edu.ph}
\newcommand{\orcidB}{0000-0002-4462-2851}
\newcommand{\institutionB}{
		National Institute of Physics, 
		University of the Philippines, Diliman, Quezon City, Philippines}

\newcommand{\authorC}{Nicko Angelo Rabang}
\newcommand{\emailC}{nrabang@nip.upd.edu.ph}
\newcommand{\orcidC}{}
\newcommand{\institutionC}{
		National Institute of Physics, 
		University of the Philippines, Diliman, Quezon City, Philippines}


\begin{center}
	\textbf{\authorC}$^{a}$,
    \textbf{\authorA}$^{b}$,
	and
	\textbf{\authorB}$^{a}$
	\par

	\texttt{\footnotesize \email{\emailC}}, 
	\texttt{\footnotesize \email{\emailA}}~\orcidlink{\orcidA},
    \texttt{\footnotesize \email{\emailB}}~\orcidlink{\orcidB}
\end{center}

\begin{quotation}\noindent
	\footnotesize
	\noindent$^{a}$
	\textit{\institutionB} 
	\\ $^{b}$ \textit{\institutionA} 
	\\ 
	{\phantom{.}\;
		\scriptsize{
		Authors listed in the order required by University of the Philippines: Ph.D student first, advisor last.
		}
	}
\end{quotation}

\begin{abstract}
\noindent 
A secluded, massive Abelian gauge boson called a dark $Z'$ may interact with the Standard Model through kinetic mixing and mass mixing in the Higgs sector. 
We determine the sensitivity of a future high-energy muon collider to discover such a particle and determine its mixing parameters. 
We examine a dark $Z'$ with mass from 100~GeV up to the collider energy for a set of collider benchmarks up to 14~TeV.
We show the discovery reach and compare to the current and proposed future colliders.
A muon collider is sensitive to two complementary production modes: radiative return (muon fusion with an associated  photon), and vector boson fusion of $W$ bosons. An observable photon distinguishes these production modes and the relative rates of these processes allows one to determine the relative mixing. Soft and collinear photons in the radiative return diagram contribute to the same final state as vector boson fusion.
We show that these relative rates alone can determine the mixing to an accuracy comparable to that of a fully polarized muon beam using a left--right asymmetry.
\end{abstract}

\newpage
\small
\setcounter{tocdepth}{2}
\tableofcontents
\normalsize
\clearpage


\section{Introduction}

One of the bold proposals for the future of particle physics is a muon collider.
Such a machine would be a powerful tool to search for extensions of the electroweak sector of the Standard Model.
We examine the reach of a muon collider to elucidate a particular class of new physics: models with a hidden (dark) Abelian gauge field $Z'$. These are called dark $Z'$s when they are heavier than the electroweak scale.\footnote{%
    These particles are known as dark photons when they are much lighter than the electroweak scale.} %
It is well known that these fields may interact with the Standard Model particles through kinetic mixing with hypercharge.
In fact, the dark $Z'$ may have a second portal interaction, a mass mixing term with the $Z$ boson due to an extended Higgs sector.
We investigate the phenomenology of dark $Z'$ produced via radiative return and vector boson fusion (\VBF{}) channels at a future high-energy muon collider. We highlight the unique role of the latter process to distinguish dark $Z'$ models with mass mixing even without polarized beams. This is because kinetic mixing enhances the interactions between the Goldstone bosons associated with the $W^\pm$ and $Z'$ bosons. A muon collider can thus determine the dark $Z'$ couplings in a way that would otherwise require a precision measurement program at a linear collider with polarized beams.
We focus on dark $Z'$ masses from 100~GeV up to the muon collider energy (up to 14~TeV).

\paragraph{Earlier Work}
This study complements earlier work on the same model at an up-to-500~GeV linear collider~\cite{San:2022uud}.
The proposal of kinetic mixing has a nearly 70-year history~\cite{Kobzarev:1966qya,Okun:1982xi,Holdom:1985ag,Holdom:1986eq}; we refer to Ref.~\cite{Fabbrichesi:2020wbt} for a review.
The role of mass mixing in an extended Higgs sector was presented in Refs.~\cite{Foot:1991kb,Davoudiasl:2012ag, Davoudiasl:2013aya,Gopalakrishna:2008dv}.\footnote{These references introduce alternative UV completions for the low-energy dark $Z'$ scenario. We focus on the model in Ref.~\cite{San:2022uud} because it cleanly separates the dimensionful parameters that control the dark $Z'$ mass and mixing.} The rich structure of the combined mass and kinetic mixing model was recently highlighted in an analysis of electroweak precision observables that show strong constraints on mass mixing and the extended scalar sector that produces it~\cite{Bertuzzo:2025ejw}.\footnote{Ref.~\cite{Bertuzzo:2025ejw} also introduces an elegant alternative parameterization of the theory in terms of mixing angles. We instead work to leading order in these small mixing parameters. We briefly connect to their parameterization in Appendix~\ref{sec:UVmode:EffParams}.} 
We focus on heavier dark $Z'$ masses in the decoupling limit of the scalar sector to highlight the opportunities of a muon collider on the low-energy dark $Z'$ theory. Earlier work on the phenomenology of this model below the electroweak scale may be found in, e.g.~Ref.~\cite{Curtin:2014cca}.
Recent work on $Z'$ models in a similar mass range at muon have focused on the sensitivity to visible and invisible channels~\cite{Barik:2024kwv,Airen:2024iiy} and its modifications to the parton distribution functions of the muon beam~\cite{Asadi:2026kpt}.

The primary bounds on the dark $Z'$ model in the $m_{Z'}\gtrsim 100~\text{GeV}$ regime comes from Drell--Yan production at the Large Hadron Collider (\acro{LHC})~\cite{Davoudiasl:2012ag, Davoudiasl:2013aya, Elkafrawy:2021mrm, Cheng:2024hvq, Sun:2023kfu}. 
Complementary information about the chiral couplings---and therefore the relative kinetic versus mass mixing---may be gleaned from polarized beam studies at a linear collider~\cite{San:2022uud}.
Studies of extended Abelian sectors at future electron--positron colliders include Refs.~\cite{Karliner:2015tga, He:2017zzr}, which highlighted the role of the \emph{radiative return} process\footnote{This is also known as the recoil mass technique.} where an initial state lepton radiates hard photon so that the $e^+e^- \to Z'$ subprocess may produce the new state on resonance. The hard photon becomes an additional handle to tag these events. In this paper we contrast this to the $W^+W^-\to Z'$ vector boson fusion process where the $W$s are emitted from the muon beam; this does not produce a hard photon and is so distinguishable from the radiative return channel. Because \VBF{} encodes the interactions of the eaten Goldstone modes, it is especially sensitive to the mass mixing parameter.

Vector boson fusion is, in fact, one of the key motivations for a future muon collider~\cite{Costantini:2020stv,AlAli:2021let, Franceschini:2021aqd, Casarsa:2022iod, Black:2022cth,Roser:2022sht, MuonCollider:2022xlm, Accettura:2023ked,InternationalMuonCollider:2024jyv}. It is a powerful probe of the electroweak sector that is otherwise not accessible in current or other planned future experiments~\cite{Li:2025ptq,Han:2020pif,Chiesa:2021qpr}, see e.g.~\cite{Costantini:2020stv} for a review and early studies of Higgs studies at a future lepton collider~\cite{Dawson:1984ta, Hikasa:1985ee, Altarelli:1987ue, Kilian:1995tr, Gunion:1998jc}.
This study continues an exploration of the role of \VBF{} to search for new physics at a muon collider~\cite{Bao:2022onq,Braathen:2024ckk, Bandyopadhyay:2024gyg, Bandyopadhyay:2024plc, Dehghani:2025xkd,  Cao:2025fla, Kwok:2023dck, Acanfora:2026aay}. Ref.~\cite{Leike:1998wr} reviews the phenomenology of additional neutral gauge bosons with some discussion of muon colliders.
In our comparison to radiative return, we use mass-dependent cuts following the optimal recoil mass technique for heavy dilepton resonances~\cite{Cheung:2025uaz}. Ref.~\cite{Korshynska:2024suh} highlights the role of polarized muon collider beams discriminate between different heavy gauge boson models based on their chiral couplings. We follow their methodology to examine the coupling determination of a 30\% polarized muon collider---we find that the \VBF{} mode for an unpolarized beam is already more powerful for the dark $Z'$ model. 

We calculate the \VBF{} process with the full, tree-level matrix element including the final state neutrinos. This is a sufficient and conservative estimate for our level of phenomenological study motivating a future collider. However, we emphasize the a long history of examining the partonic composition of colliding high-energy lepton beams~\cite{Dawson:1984gx, Kane:1984bb, vonWeizsacker:1934nji, Williams:1934ad} has led to ongoing work in the effective vector boson approximation~\cite{Ruiz:2021tdt} that is especially timely for muon collider studies.

\paragraph{This paper is organized as follows}
We present the dark $Z'$ model in Section~\ref{sec:darkZ:EFT}.
Section~\ref{sec:dark:Z:at:muon:collider} introduces the properties of the radiative return and \VBF{} production modes that are the focus of this study. 
We discuss the Monte Carlo simulation framework and phenomenological analysis in Section~\ref{sec: Search strategy}. 
In Section~\ref{sec:Results} we present our results as sensitivity projections of the proposed muon collider, comparisons to other collider facilities, and as a $\chi^2$-test analysis to determine the mixing parameters. Summarize our findings in Section~\ref{sec:Conclusion}. The appendices present additional model details and a set of analytic partial decay widths.

\section{Dark Z' Effective Theory}
\label{sec:darkZ:EFT}
 
\subsection{Model Parameters}

We use a benchmark model for an Abelian dark $Z'$ gauge boson with both kinetic and mass mixing~\cite{San:2022uud}. Three parameters characterize the effective theory near the scale of the $Z'$ mass:
\begin{enumerate}
    \item Kinetic mixing, $\varepsilon$, between the gauge eigenstate $Z'$ and the hypercharge boson,
    \item Mass mixing, $\kappa$, between the gauge eigenstate $Z'$ and the neutral SU(2)$_\text{L}$ gauge boson,
    \item The physical $Z'$ mass, $m_{Z'}$. In this work we focus on $m_{Z'}$ heavier than electroweak scale.
\end{enumerate}
Both mixing parameters are dimensionless and assumed to be much smaller than one. 
\subsection{Effective Theory and Leading Order Approximations}

In Appendix~\ref{app:model:summary} we review a benchmark \acro{UV}-complete realization of this model and the decoupling limit that realizes the three-parameter effective theory used in this study. The low-energy interactions of the dark $Z'$ follow from writing the electroweak hypercharge and neutral weak boson in terms of the mass eigenstate $A$, $Z$, and $Z'$,
\begin{align}
    B &= 
    \text{c}_\text{W} A - \text{s}_\text{W} [Z+(\text{s}_\delta - \varepsilon \text{c}_\text{W}\inv) Z'] 
    &
    W^3 = \text{s}_\text{W} A 
        + \text{c}_\text{W} (Z + \sin\delta Z') 
    \ ,
    \label{eq:EW:gauge:to:mass:eigenbasis}
\end{align}
where $\text{s}_\text{W}$ and $\text{c}_\text{W}$ are the sine and cosine of the Weinberg angle, and $\delta$ is the angle that diagonalizes the $Z$--$Z'$ mass term,
\begin{align} 
    \delta &= \frac{ -m^2_Z }{ m^2_{Z'} - m^2_{Z} }  \zeta
    &
    \zeta \defeq \kappa + \varepsilon \tan\theta_\text{W} \ll 1 \ .
    \label{eq:delta:zeta:definition}
\end{align}
We work to leading order in the small parameter $\zeta$, which simplifies the analysis by allowing us to (1) use the physical masses in the expression for $\delta$ and (2) subsequently take $\sin\delta \approx \delta$. 
We further take the limit where the ratio of the $Z$ to $Z'$ mass is small and we may define an effective dark $Z'$ coupling to Standard Model fermions,
\begin{align}
    \xi^2 &\defeq \frac{m_Z^2}{m_{Z'}^2}
    &
    \delta &\approx -\xi^2 \zeta
    &
    g_{Z'} \defeq \frac{g}{\text{c}_\text{W}}
    \left(\xi^2 \kappa + \varepsilon \tan^2 \theta_\text{W}\right) \ .
    \label{eq:g:Zp:definition}
\end{align}
The $\zeta$ factor in the mixing angle and the effective coupling of the dark $Z'$ to fermions $g_{Z'}$ include the effect of both types of mixing, but the latter suppresses the mass mixing by $\xi^2$.
The presence of both kinetic and mass mixing implies that the relative sign of these mixing sources is physical.

\subsection{Interactions}
We summarize the key Feynman rules for the model. 
The dark $Z'$ vector and axial couplings to fermions $f$ are proportional to the effective coupling $g_{Z'}$ in \eqref{eq:g:Zp:definition},
\begin{align}
    g_{Z'f}^\text{V} &\defeq
    \varepsilon e Q^f_\text{EM}
     - g_{Z'} Q_{Z\text{V}}^f
    &
    g_{Z'f}^\text{A} &\defeq
     - g_{Z'} Q_{Z\text{A}}^f \ ,
    \label{eq:gZp:V:A}
\\
\intertext{
where  the $Q$'s are the fermion's electric, vector, and axial $Z$ charges,
}
    Q_{Z\text{V}} 
    &\defeq
        \frac{1}{2}T^3
        - Q_\text{EM} \text{s}_\text{W}^2
    &
    Q_{Z\text{A}} 
    &\defeq
        \frac{1}{2}T^3 \ .
    \label{eq:Z:charges}
\end{align}
The Feynman rule for this interaction is:
\begin{align}
    \eqfig{\includegraphics[height=3.5em]{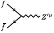}}
    &=
    i \gamma^\mu(g_{Z'f}^\text{V} + g_{Z'f}^\text{A} \gamma^5) \ . 
\end{align}

The $Z'$ interactions with bosons are independent of $g_{Z'}$ and instead are inherited from $\delta = -\xi^2\zeta$ in \eqref{eq:delta:zeta:definition}. While this introduces an additional suppression by the difference of the $Z'$ and weak scales, $\xi^2$, observe that in $\zeta$ the mass mixing is not suppressed relative to the kinetic mixing as it is in $g_{Z'}$. 
The $Z'Zh$ interaction is
\begin{align}
    \eqfig{\includegraphics[height=3.8em]{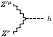}}
    &=
     i \frac{2 m^2_{Z}}{v} \xi^2\zeta g^{\mu \nu} \ .
\end{align}
There is no $Z'hh$ interaction because the $Z' (H^{\dagger} \partial_{\mu} H - H \partial_{\mu} H^{\dagger})$ term is zero for the neutral component of the Higgs doublet. 

The dark $Z'$ interacts with gauge bosons through the weak gauge boson self-interactions,
\begin{align}
    \eqfig{\includegraphics[height=3.5em]{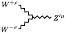}}
    &=
    i g \text{c}_{\text{W}} \xi^2 \zeta
    \left[
        g^{\rho \sigma} \left(p_1 - p_2 \right)^{\mu} 
        + g^{\sigma \mu} \left(p_2 - p_3 \right)^{\rho} 
        + g^{\mu \rho} \left(p_3 - p_1 \right)^{\sigma} 
    \right]
    \\
    \eqfig{\includegraphics[height=3.5em]{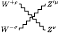}}
    &=
     i g^2 \text{c}^2_{\text{W}} \xi^2 \zeta  \left(g^{\nu \rho} g^{\mu \sigma} + g^{\nu \sigma} g^{\mu \rho} - 2 g^{\nu \mu} g^{\rho \sigma} \right) \ .
    \\
    \eqfig{\includegraphics[height=3.5em]{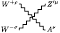}}
    &=
     i g^2 \text{c}_{\text{W}} \text{s}_{\text{W}} \xi^2 \zeta  \left(g^{\nu \rho} g^{\mu \sigma} + g^{\nu \sigma} g^{\mu \rho} - 2 g^{\nu \mu} g^{\rho \sigma} \right) \ .
\end{align}

\subsection{Role of Mass Mixing in Production and Decay}

The dark $Z'$ interactions with bosons is suppressed by the heaviness of the $Z'$ relative to the weak scale, $\xi^2$. This suppression can be mitigated in the production and decay of the $Z'$. 

\paragraph{Production} A key opportunity of a high-energy muon collider that it is effectively an electroweak gauge boson collider~\cite{Costantini:2020stv}. Dark $Z'$ particles may be produced from vector boson fusion $WW\to Z'$. The $\xi^2$ suppression of this interaction compared to muon annihilation is partially overcome by the effective $W$ parton distribution function.
At large collider center of mass energy $\sqrt{s}$, there are logarithmic enhancements for soft and collinear $W$ emission .
More significantly, because the $W$s are produced at a fraction of the beam energy, one may access $Z'$ production that is closer to threshold rather than suffering a $\hat{s}^{-1}$ scaling for cross sections at large partonic center of mass energy, $\sqrt{\hat{s}}$.

\begin{figure}[t]
    \centering
    \includegraphics[width=\linewidth]{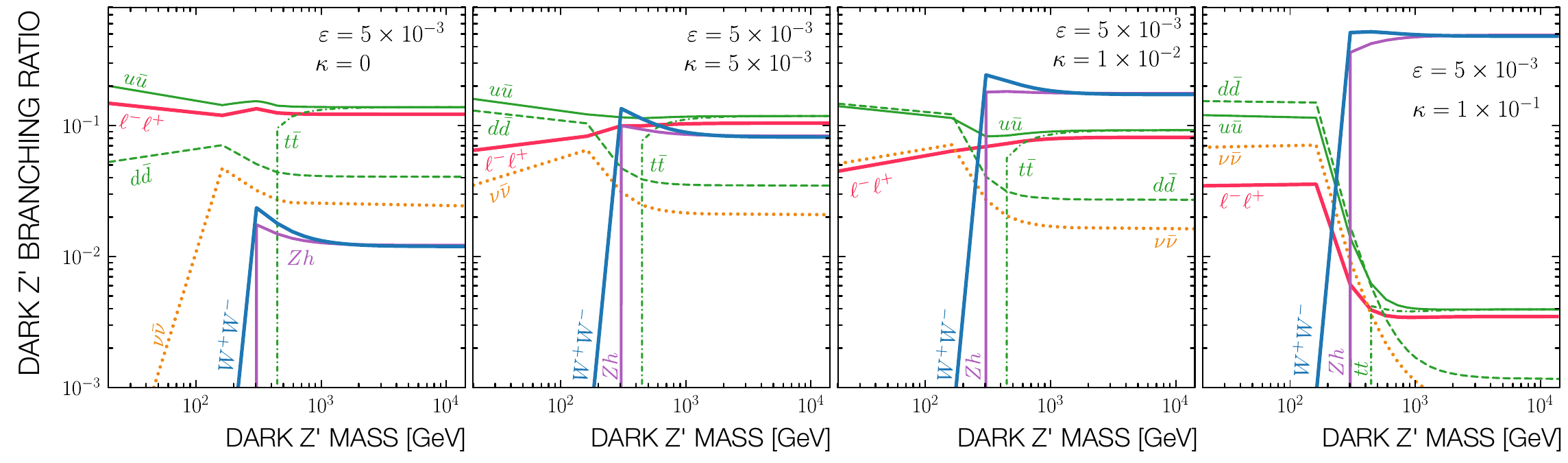}
    \caption{Branching ratio of dark $Z'$ as a function of its mass for different values of mass mixing parameter ($\kappa = 0, \; 5 \times 10^{-3}, \; 1 \times 10^{-2}, \; 1 \times 10^{-1}$) at fixed kinetic mixing parameter, $\varepsilon = 5 \times 10^{-3}$. The branching ratios are approximately constant for $m_{Z'} \gtrsim 400 \; \mathrm{GeV}$. 
    The convergence of the $Z'\to Zh$ and $Z'\to W^+W^-$ branching ratios at large $m_{Z'}$ reflects the Goldstone equivalence theorem.
    The lepton ($\ell^+\ell^-$) and neutrino $\nu\bar\nu$ lines include all three generations. The charm (strange and bottom) lines are effectively degenerate with the up (down) lines.
    }
    \label{fig:Dark:Z:Decay:Widths}
\end{figure}

\paragraph{Decay} The $Z'WW$ coupling affects the $Z'$ branching fractions for heavy dark $Z'$s. We demonstrate the contribution of mass mixing in Figure~\ref{fig:Dark:Z:Decay:Widths}.
Unlike the fermionic couplings where the mass mixing contribution $\kappa$ is suppressed by $\xi^2$ relative to kinetic mixing $\varepsilon$, the coupling of $Z'$s to bosons receive contributions from $\kappa$ and $\varepsilon$ to the same order in $\xi^2$. On the other hand, these bosonic interactions scale like $\xi^2$, and so one might expect the kinetic mixing contribution to fermionic final states to dominate.

As highlighted in Ref.~\cite{San:2022uud}, this $\xi^2$ suppression is overwhelmed for large $m_{Z'}$. One may see this in the spin sum for each outgoing $W$ which contributes a $k_\mu k_\nu/m_W^2\sim \xi^{-2}$ enhancement. Figure~\ref{fig:Dark:Z:Decay:Widths} verifies that the $WW$ and $Zh$ final state branching ratios approach each other as expected from the Goldstone boson equivalence theorem.

\subsection{Bounds}

Sub-GeV dark $Z'$ particles are constrained by the anomalous magnetic moments of leptons, rare meson decays, and parity violation~\cite{Davoudiasl:2012ag, Davoudiasl:2012qa, Davoudiasl:2014kua}. 
For energies below the $Z$ mass one may recast the bounds on dark photons~\cite{Alexander:2016aln,Battaglieri:2017aum,Fabbrichesi:2020wbt,Caputo:2021eaa}. For $m_{Z'}\approx m_Z$, the mixing angle \eqref{eq:delta:zeta:definition} is maximized\footnote{The approximation in \eqref{eq:delta:zeta:definition} breaks down in this regime; a full treatment gives $\tan(2\delta)\sim (m_{Z'}-m_Z)^{-1}$, as is familiar in other examples of particle mixing in the Standard Model.} and the model is tightly constrained by the way this mixing modifies the electroweak sector. Refs.~\cite{Curtin:2014cca,San:2022uud} took initial steps towards quantifying these bounds, but a recent analysis in Ref.~\cite{Bertuzzo:2025ejw} gives a systematic application of electroweak precision observables to the dark $Z'$ scenario. We provide a dictionary between our model parameters in \eqref{eq:Bertuzzo:comparison}. 

Instead, we focus primarily on dark $Z'$ masses heavier than the weak scale. In this regime, dark $Z'$s are produced by Drell--Yan production at \acro{LHC} and can be constrained by searches for dilepton resonances \cite{CMS:2019buh, CMS:2021ctt, ATLAS:2019erb}. Ref.~\cite{San:2022uud} compared the present reach of the \acro{LHC} and the anticipated \acro{HL-LHC} program to a hypothetical electron collider (\acro{ILC}) running at up to $\sqrt{s}=500~\text{GeV}$. Their recast of Drell--Yan production gives an upper bound on the couplings
\begin{align}
    \xi^2\kappa +\varepsilon\tan^2\theta_\text{W}
    &\lesssim \mathcal O(10^{-3})
    &
    200~\text{GeV} \lesssim m_{Z'}\lesssim 500~\text{GeV}
\end{align}
with \acro{ILC} reach about an order of magnitude stronger. The Drell--Yan bounds extend to dark $Z'$ masses of a few TeV with the bound on the mixing parameters softened by a factor of few.
In this energy range, both the \acro{ILC} and \acro{LHC} produce dark $Z'$ bosons through their coupling to fermions. For this coupling, the dependence on the mass mixing $\kappa$ is suppressed as $m_{Z'}$ becomes heavy. One of the key features of a high-energy muon collider is the re-emergence of $\kappa$'s significance in additional production channels and its decays.

\begin{figure}[t]
    \centering
    \includegraphics[width=0.9\textwidth]{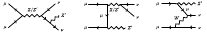}
    \caption{Representative non-VBF diagrams for the process $\mu^- \mu^+ \rightarrow Z' \nu_{\mu} \bar{\nu}_{\mu}$.}
    \label{fig:Non_VBS_Diagrams}
\end{figure}

\section{%
    \texorpdfstring{%
    Dark $Z'$ at a Muon Collider%
    }{%
    Dark Z' at a Muon Collider%
    }} \label{sec:dark:Z:at:muon:collider}

\subsection{Collider Benchmarks}
\label{sec:collider:benchmarks}
A standard benchmark scenario is a 10~TeV muon collider with $10$~ab$^{-1}$ of integrated luminosity for which one expects percent level precision for Standard Model electroweak processes~\cite{Accettura:2023ked}. This machine may be preceded by a 3~TeV low-energy phase. We thus examine the following benchmark collider energies and integrated luminosities: 
\begin{itemize}
    \item $\sqrt{s} = 3~\text{TeV}$ with $\mathcal L = 1$~ab$^{-1}$ integrated luminosity
    \item $\sqrt{s} = 10~\text{TeV}$ with $\mathcal L = 10$~ab$^{-1}$  integrated luminosity
    \item $\sqrt{s} = 14~\text{TeV}$ with $\mathcal L = 20$~ab$^{-1}$  integrated luminosity \ .
\end{itemize}
The International Muon Collider Collaboration identify these benchmarks as having comparable physics reach to a 100 TeV Future Circular Collider~\cite{Casarsa:2023vqx, Delahaye:2019omf, Long:2020wfp}.
Each benchmark carries a different integrated luminosity. This is because unlike other types of high-energy colliders, the instantaneous luminosity of a muon collider in fact scales with the square of the center-of-mass energy, $s$, due to time dilation of the muon decays and the reduced beamstrahlung that permits tighter muon bunches at high energies~\cite{Accettura:2023ked}\footnote{This luminosity scaling with $s$ is one of the highlights of the muon collider proposal. A key challenge of high-energy colliders is that signal cross sections typically scale like the inverse of the squared center of mass energy $s^{-1}$. For more energetic $s$, one thus requires a correspondingly brighter luminosity for the same precision of signal events.}.

\subsection{Effective Vector Boson Approximation \& Sudakov Uncertainties}
\label{sec:on:EVA}

At high energies, collinear electroweak gauge boson emission is logarithmically enhanced and one may treat the muon beam as having a distribution of $W$ partons. This is a key feature of the Standard Model program of a muon collider it opens up the vector boson fusion (\VBF{}) channel for Higgs production.\footnote{From the Goldstone boson equivalence theorem, the dark $Z'$ longitudinal mode mixes with the neutral Goldstone boson in the Higgs doublet. We thus expect \VBF{} to be a significant production mode for dark $Z'$ bosons with large mixing. }
%
%
The effective vector boson approximation (\acro{EVA})~\cite{Ruiz:2021tdt, Han:2020uid, Bigaran:2025rvb} encompasses the massless Weizs\"{a}cker-Williams~\cite{vonWeizsacker:1934nji, Williams:1934ad} and massive effective $W/Z$ approximations~\cite{Dawson:1984gx, Kane:1984bb} under a fully polarized framework.\footnote{The effective $W/Z$ approximation, in turn, extends the the Weizs\"{a}cker-Williams approximation to include the emission of massive gauge bosons~\cite{vonWeizsacker:1934nji, Williams:1934ad}. The kinematic regime where the EVA is a good approximation of full matrix elements has recently been revisited and clarified in e.g.~\cite[\S 3.6]{Dahlen:2025udl}.}
The current implementation captures the leading logarithm corrections,
$
\mathcal O
        \left( \alpha_\text{W}
         \ln \mu^2_\text{f} / m^2_W \right)
$, 
 of the perturbative expansion,
where 
    $\alpha_\text{W}$ is the weak coupling constant, 
    $m_W$ is the mass of the $W$ boson, 
    and $\mu_\text{f}\sim m_{Z'}$ is the factorization scale. 

Applications of \acro{EVA} for muon beams focuses processes with a muon emitting a single vector boson that scatters off an electron~\cite{Ruiz:2021tdt, Bigaran:2025rvb}. This mixed-flavor initial state removes diagrams that would contribute to the matrix element but are not included in \acro{EVA}.
We cannot do this for signal processes based on vector boson fusion $\mu^- \mu^+ \rightarrow Z' \nu_{\mu} \bar{\nu}_{\mu}$;  additional non-\acro{VBF} Feynman diagrams like those in Figure~\ref{fig:Non_VBS_Diagrams} interfere with the \VBF{} diagrams in \acro{EVA}. 

For this reason, we do \emph{not} use \acro{EVA} and instead 
calculate the full tree-level matrix elements for \VBF{} processes which includes the potential interference of non-VBF diagrams. 
In this estimate, the Monte Carlo simulation populates positive contributions of phase space of $\mu \to \nu W$ real emission logarithms, but does not account for the negative contributions of virtual (loop) diagrams:
\begin{align} 
\label{eq:EVA}
    \sigma 
    &\approx 
    \sigma_\text{MC} 
    \left[
        1 
        -
        \mathcal O
        \left(
            \frac{ \alpha_\text{W} }{ 2 \pi }
         \ln \frac{ m_{Z'}^2 }{ m^2_W }
        \right)
        - 
        \mathcal O
        \left(
            \frac{ \alpha_\text{W} }{ 2 \pi }
         \ln^2 \frac{ m_{Z'}^2 }{ m^2_W }
        \right)
    \right]
\end{align}
where $\sigma_\text{MC}$ is the tree-level Monte Carlo cross section and we identify the factorization scale with that of the hard process.
The denominator of the corrections is $2\pi$ rather than $4\pi$ to account for the two $W$ emissions for \acro{VBF}. 
The first correction is a percent-level correction over the entire range of heavy $m_{Z'}\lesssim \sqrt{s}$ we consider. 
The second correction is the Sudakov double logarithm and is the dominant source of error. For a $m_{Z'} = 700$~GeV dark $Z'$, this correction is order 10\%. The correction rises to 25\% for a heavier $m_{Z'}= 2.5$~TeV boson. 
Because the goal of this paper is to demonstrate the phenomenological interplay of \VBF{} and radiative return rather than detailed targets, we keep this level of accuracy and leave more detailed predictions to future work with ongoing \acro{EVA} improvements.

\begin{figure}[t]
    \centering
    \includegraphics[width=\textwidth]{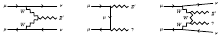}
    \caption{Representative Feynman diagrams for the \VBF{} contribution to $\mu^- \mu^+ \rightarrow Z' \nu_{\mu} \bar{\nu}_{\mu}$ \textsc{(left panel)} and the radiative return channel $\mu^- \mu^+ \rightarrow Z' \gamma$ \textsc{(middle panel)}. On the right is an example of the combined VBF--radiative return processes that we do not include in our analysis because they are always subdominant.}
    \label{fig:Feynman_Diagrams}
\end{figure}

\subsection{%
\texorpdfstring{%
    Dark $Z'$ Production Modes%
    }{%
    Dark Z' Production Modes%
    }
}
\label{sec:Zp:production:modes}

We examine complementary dark $Z'$ production modes at a muon collider: radiative return and vector boson fusion (\VBF{}). We assume that the dark $Z'$ decay width is small so that the narrow width approximation is valid.\footnote{See Ref.~\cite{San:2022uud} for a discussion of the possible experimental challenges when searching for such a narrow resonance at a linear collider.} We show characteristic diagrams in Figure~\ref{fig:Feynman_Diagrams}. These production modes approach the dark $Z'$ production resonance even when the muon beam collision energy $\sqrt{s}$ is much larger than $m_{Z'}$.

\paragraph{Radiative Return}
Radiative return~\cite{Czyz:2002np, Binner:1999bt, Melnikov:2000gs, Denner:2000jv, Spagnolo:1998mt} is the process $\ell^+\ell^-\to Z'\gamma$ by which one of the incoming leptons emits a hard photon to carry away the `excess' energy so that the $Z'$ may be produced on shell~\cite{Cheung:2025uaz, Chakrabarty:2014pja, Draper:2018ljh, Han:2020pif, Zhu:2022lzv, Dasgupta:2023zrh, Chen:2022yiu, Denizli:2023rqe, Forslund:2023reu, Li:2024joa, Barik:2024kwv, Cipressi:2026wtb}. 
The $2\to 1$ process that is forbidden by momentum conservation is replaced by a new kinematic configuration 
at the cost of $\alpha/4\pi$.
The emitted photon energy is
\begin{align}
    E_{\gamma} &= \frac{s - m^2_{Z'}}{2 \sqrt{s}} \ .
    \label{eq:radiative:return:photon:energy}
\end{align}

\begin{figure}[t]
    \centering
    \includegraphics[width=\linewidth]{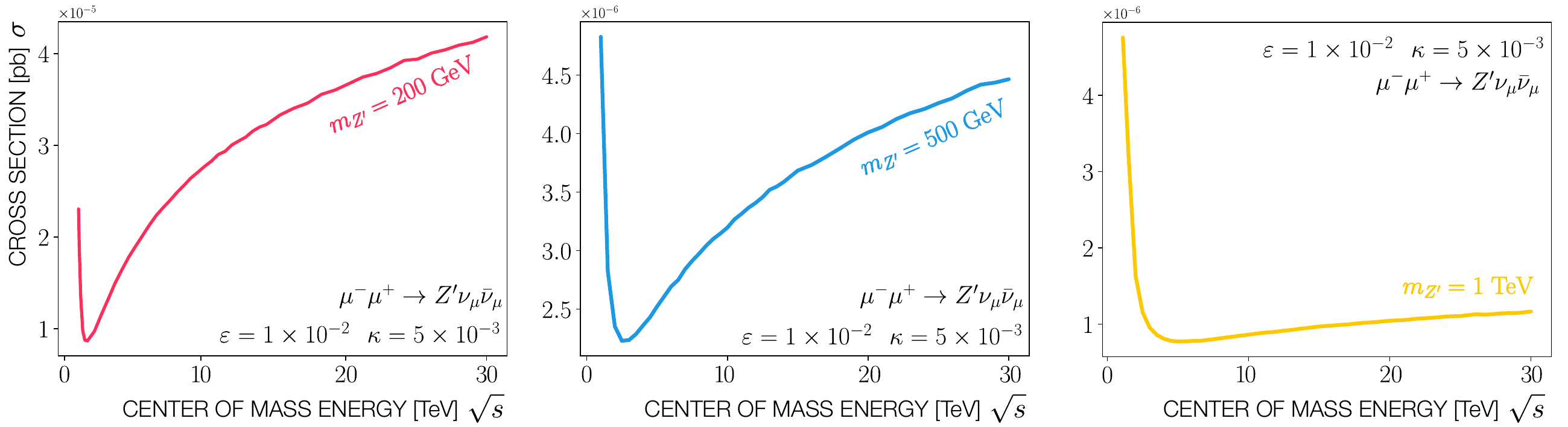}
    \caption{%
    Vector boson fusion cross section as a function of the collider energy corresponding to the process for three values of the dark $Z'$ mass and benchmark mixing parameters.}
    \label{fig:VBF:xsec}
\end{figure}

\paragraph{Vector Boson Fusion}
Vector boson fusion (\VBF{}) constitutes a key opportunity of the muon collider electroweak physics program~\cite{AlAli:2021let, Franceschini:2021aqd, Casarsa:2022iod, Black:2022cth,Roser:2022sht, MuonCollider:2022xlm, Accettura:2023ked,InternationalMuonCollider:2024jyv} and offers a promising avenue for new physics searches at multi-TeV muon colliders~\cite{Dehghani:2025xkd, Li:2025ptq, Cao:2025fla, Liu:2023jta, Braathen:2024ckk, Costantini:2020stv, Buttazzo:2018qqp, Han:2023njx, Bandyopadhyay:2024gyg, deLima:2025ctj,Bandyopadhyay:2024plc}. 
At center-of-mass energies above the electroweak scale, nearly on-shell $W$ bosons may be emitted from the muon beam and effectively behave as partons. These $W$s may fuse into a dark $Z'$ by $W^+W^- \to Z'$. As described in Section~\ref{sec:on:EVA}, we treat this process with a full tree-level matrix element rather than in the effective vector boson approximation.

In Figure~\ref{fig:VBF:xsec} we plot the $\mu^- \mu^+ \rightarrow Z' \nu_{\mu} \bar{\nu}_{\mu}$ cross section as a function of the center of mass energy $\sqrt{s}$ for benchmark values of dark $Z'$ mass and mixings. We confirm that apart from the expected $s^{-1}$ scaling at low energies, the cross sections indeed scale with the expected $\log^2(s/m_W^2)$ coming from the collinear enhancement from the production of $W$ bosons with a small fraction of the beam energy.

\paragraph{%
\texorpdfstring{Radiative Return Mimicking \VBF{} ($\mu$\acro{PDF})}%
{Radiative Return Mimicking VBF (muPDF)}
}
When the photon in a radiative return topology is soft or collinear, the cross section picks up logarithmic enhancements analogous to those in Section~\ref{sec:on:EVA}. There is a large contribution from these events and they are best treated separately as muon annihilation, $\mu^+\mu^-\to Z'$, from a muon parton distribution function within the muon beam whose factorization scale $\mu_\text{F}$ is matched to the photon $p_\text{T}$ cut. While these events are excluded from the radiative return signature, they contribute to the same dielectron/no photon final state as \acro{VBF}. Appendix~\ref{app:RR:contribution:to:VBF} describes the qualitative distribution of these events.

\paragraph{Production Cross Sections}

\begin{figure}[t]
    \centering
    \includegraphics[width=\linewidth]{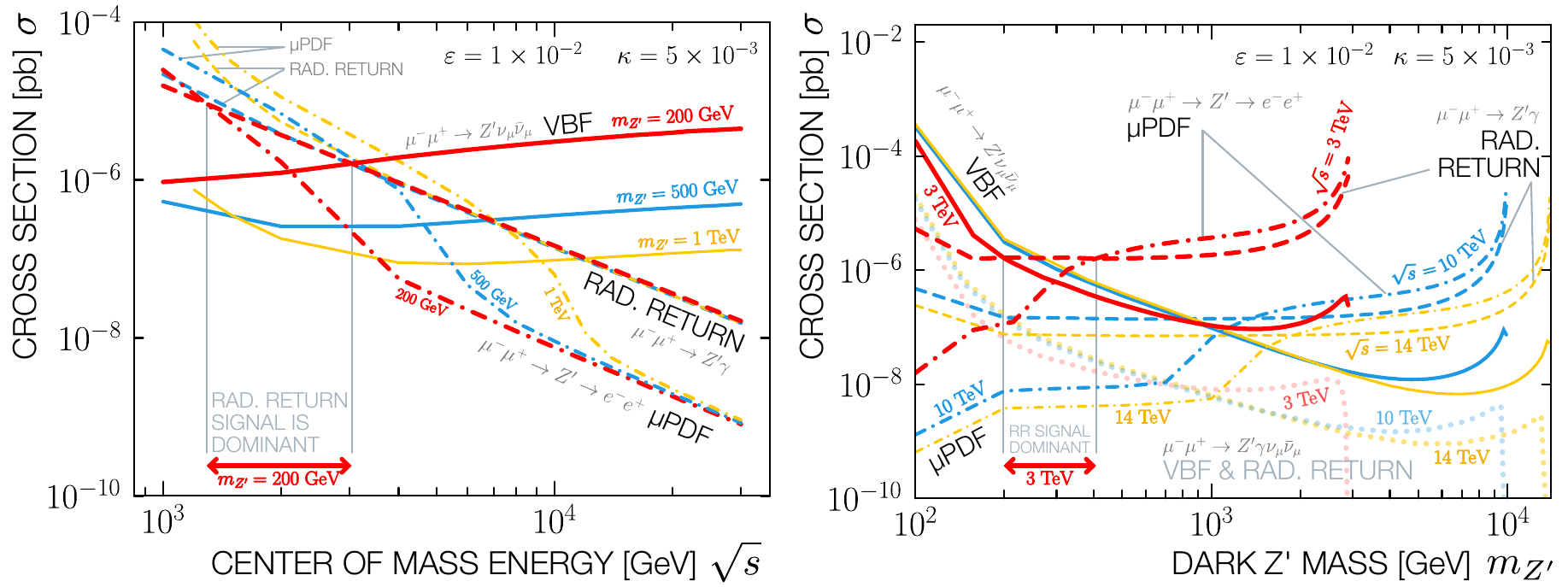}
    \caption{%
        Parton-level dark $Z'$ production cross sections for radiative return (dashed) and vector boson fusion (solid) using methods in Section~\ref{sec:Monte:Carlo:Simulation}.
        We separate the $\mu$\acro{PDF} cross section (dot--dashed) for radiative return topologies with a soft/collinear photon---these contribute to the \VBF{} signal.
        In each plot we highlight the range where the radiative return (dielectron plus visible photon) signal dominates for one of the masses/collider energies.
        \textsc{Left}: cross section as a function of collider energy for representative $Z'$ masses (colors). The radiative return lines overlap for $\sqrt{s}\gtrsim 3~\text{TeV}$. 
        \textsc{Right}: cross section as a function of $Z'$ mass for representative collider energies (colors). Dotted lines correspond to \VBF{} with a hard photon, $\mu^- \mu^+ \rightarrow Z' \gamma \nu_{\mu} \bar{\nu}_{\mu}$,
        which is always subdominant.
        }
    \label{fig:Cross Sections}
\end{figure}

Figure~\ref{fig:Cross Sections} shows the parton-level cross sections for dark $Z'$ production via radiative return and vector boson fusion as a function of center of mass energy and of the dark $Z'$ mass. 
We separate the muon $\mu$\acro{PDF} cross section which corresponds to radiative return diagrams whose photon is too soft or collinear to be observable.

The radiative return cross section demonstrates the characteristic $s^{-1}$ scaling away from the $m_{Z'}$ resonance.
    On the other hand, the \VBF{} cross section carries logarithmic enhancements in $s/m_{W}^2$ from real emissions in the phase space integral. 
    The $\mu$\acro{PDF} cross section similarly is similarly enhanced in regions where $m_{Z'}$ is close to $\sqrt{s}$.
    This leads to different regimes where each production mode dominates. Of course, \VBF{} and $\mu$\acro{PDF} production modes produce the same signal. Figure~\ref{fig:Cross Sections} highlights the regime where the radiative return signal dominates; to either side of this region the dielectron/no photon signal dominates, though these are due to different underlying dark $Z'$ production mechanisms.
 
    The true \VBF{} cross section dominates when the dark $Z'$ mass is light compared to the collider energy. In this regime the $W$ emission is soft compared to the beam energy and picks up logarithmic enhancements.  In the opposite limit, the $\mu$\acro{PDF} dominates because the emitted unobservable photon is soft and picks up logarithmic enhancements. In between these regimes the radiative return cross section dominates; it is here that the strongest search comes from requiring the associated photon.

The crossover between the \VBF{} and radiative return regimes is seen through their scaling with collider energy,
    \begin{align}
        \frac{\sigma_{\text{VBF}}}{\sigma_\text{RR}} 
        \sim 
            \frac{s}{m^2_{Z'}} 
        \log^2\left(\frac{s}{m^2_W} \right)
        \log\left(\frac{s}{m^2_{Z'}} \right) \ .
        \label{eq:log:log:log}
    \end{align}
This is adapted from the ratio of the \VBF{} cross section to the leading-order muon annihilation cross section far above threshold~\cite{Costantini:2020stv}. The three logarithmic factors account for the collinear enhancements for each beam and a single soft enhancement.\footnote{That there is only a single soft enhancement is similar to the discussion leading to the $x_1^{-1}$ factor in \eqref{eq:PDFs:one:over:x1} in Appendix~\ref{app:RR:contribution:to:VBF}.} We replace the annihilation cross section with the radiative return cross section since both carry the same $s^{-1}$ scaling. Our results are consistent with a proportionality constant of $\mathcal O(1.5\times 10^{-5})$ with percent-level precision.
The analogous crossover between the radiative return and $\mu$\acro{PDF} regimes is approximately, 
\begin{align}
        \frac{\sigma_{ \mu\text{PDF}} }{ \sigma_{\text{RR}} } 
        \sim 
            \frac{s}{m^2_{Z'}} 
        \log\left(\frac{s}{m^2_\mu} \right)
        \ .
        \label{eq:muPDF:to:RR}
    \end{align}
This modifies replaces the logarithms in \eqref{eq:log:log:log} with a single logarithm for soft photon emission in the  the $\mu$\acro{PDF} limit of the radiative return diagram. Our results are consistent with a proportionality constant of $\mathcal O(5\times 10^{-4})$ with percent-level precision.
Both \eqref{eq:log:log:log} and \eqref{eq:muPDF:to:RR} are meant only to demonstrate the limiting scaling behavior that justify the regimes in Figure~\ref{fig:Cross Sections}.\footnote{
For example, in Figure~\ref{fig:kinematic:features:and:resolutions} the $\mu$\acro{PDF} events contain both on-shell and off-shell contributions for our benchmark models, so the far-above threshold approximation breaks down.}

For fixed collider energy, the radiative return production cross section is flat for most of its range of $m_{Z'}$ before growing when dark $Z'$ mass approaches the collider energy, $\sqrt{s}$.
    This upturn is not due to a resonance; the radiative return process ensures that the $Z'$ is always produced resonantly by adjusting the emitted photon energy.
    Instead, this growth is because the photon becomes soft in this limit and the production cross section approaches the soft singularity that is otherwise encoded in the $\mu$\acro{PDF} rate. The process effectively becomes $2\to 1$ rather than $2\to 2$.
    Indeed, the $\mu$\acro{PDF} rate follows the features of the radiative return cross section in the $m_{Z'} \to \sqrt{s}$ limit.
    Similarly, the \VBF{} production cross section grows in this limit because of soft $Z$ production (middle diagram in Figure~\ref{fig:Non_VBS_Diagrams}) and the process effectively becomes $2\to 1$ rather than $2\to 3$.

    On the right of Figure~\ref{fig:Cross Sections} we also show the combined \acro{VBF}--radiative return channel, $\mu^+\mu^- \to Z' \gamma \nu \bar\nu$. This channel has a larger cross section than pure radiative return for low $m_{Z'}$ wherein the hard photon carries away energy to help the $W$ bosons reach the $Z'$ resonance with a smaller momentum fraction of the beam. This channel is always subdominant to \VBF{} by a factor of $\mathcal O(\alpha_\text{EM})$ and so we may safely ignore it in our analysis. The same argument follows for $\mu$\acro{PDF} with an additional hard photon.

\section{Search Strategy and Methodology}
\label{sec: Search strategy}

\subsection{Signal, Background, and Preselection Cuts}
\label{sec:sig:bg:pre}

\paragraph{Decay Mode}
We focus on the case when the dark $Z'$ decays into electrons, $Z' \rightarrow e^- e^+$. The branching ratio for this decay is $\mathcal O(10^{-1})$ for typical values of the mass mixing $\kappa \lesssim \mathcal O(10^{-2})$, see Figure~\ref{fig:Dark:Z:Decay:Widths}. The dielectron channel is a clean detector benchmark that avoids large $t$-channel backgrounds for muonic final states. For large mass mixing $\kappa \gtrsim \mathcal O(10^{-1})$ and dark $Z'$s heavier than the electroweak scale the dielectron branching ratio is sub-percent and dedicated search should focus on bosonic final states---we leave this limit to future work.

\paragraph{Signal and Production}
We use the terms \emph{radiative return} and \emph{vector boson fusion} (\VBF{}) to refer to both the production channel in Section~\ref{sec:Zp:production:modes} and the collider signature. For radiative return, we assume a final state with $Z'\to e^+e^-$ and a hard photon. For \acro{VBF}, we focus only on the $Z'\to e^+e^-$ and assume neglect the neutrino missing energy that is anyway predominantly along the beamline. In each case, the invariant mass of the observed dielectron final state is used to reconstruct the dark $Z'$ resonance. For radiative return the photon energy is a complementary source of information whose utility has been highlighted in Ref.~\cite{Cheung:2025uaz} as the optimized recoil mass technique.
Radiative return events where the photon is unobservable are counted as \VBF{} events; we model these using a muon \acro{PDF} within the muon beam.

We do not consider the `VBF and radiative return' $\mu^+\mu^- \to Z'\gamma \nu\bar\nu$ channel that combines $WW\to Z'$ production with the emission of a hard photon. As shown in Figure~\ref{fig:Cross Sections}, this channel is subdominant at all $Z'$ masses considered here.

\paragraph{Background Processes} 
The Standard Model backgrounds include  $s$-channel $Z$ and $\gamma$ production of dielectron pairs and $W^+W^-$ pair production with each decaying into an electron--neutrino pair; these are decorated with an additional photon for the radiative return channel. Diagrams with $\tau^+\tau^-$ production with subsequent $\tau$ decay into an electron and two neutrinos are at least an order of magnitude smaller than the other backgrounds and are not included in this study.

In the dielectron invariant mass range we consider the radiative return cross section is largely flat and featureless with a slight upturn for large invariant masses where the final state photon is soft.
For the vector boson fusion channel, the dominant background arises from the process $\mu^- \mu^+ \rightarrow e^- e^+ \nu \bar{\nu}$. Contributions where a $Z$ or photon is produced via $W^- W^+$ fusion become dominant at high energies which are characterized by low-$p_T$ $W$ bosons. This results to a peak at low $m_{e^- e^+}$ distribution and falls off at higher values. The $s$-channel annihilation of muons into an electron--positron pair also constitutes as an additional background where whose shape tracks the muon-in-a-muon beam parton distribution function.

\subsection{Monte Carlo Simulation and Event Selection}
\label{sec:Monte:Carlo:Simulation}

We implement the dark $Z'$ model in \texttt{FeynRules}~\cite{Christensen:2008py, Alloul:2013bka} to generate a Universal Feynman Output file \cite{Degrande:2011ua}. 
We generate $1 \times 10^6$ background and $5 \times 10^{4}$ signal events using \texttt{MadGraph5aMC@NLO-v3.5.7} \cite{Alwall:2007st, Alwall:2011uj, Alwall:2014hca} subject to the following generator-level preselection cuts:
\begin{itemize}
    \item Charged leptons must lie within the detector acceptance, $|\eta^{\ell^{\pm}}| < 2.5$, with a minimum transverse momenta of $p^{\ell^{\pm}}_\text{T}>50~\text{GeV}$. 
    \item Photons must have a minimum transverse momentum of $p^{\gamma}_\text{T}>25~\text{GeV}$ and $|\eta^{\gamma}| < 2.5$ with an isolation requirement of $\Delta R_{\ell^{\pm} \gamma} > 0.4$. 
\end{itemize}
The choice of a large $p_\text{T}$ cut rejects most of the beam-induced background at a muon collider~\cite{Bandyopadhyay:2024gyg, Cheung:2025uaz, Bartosik:2019dzq}. To simulate the case of unobservable soft/collinear photons, we use the \texttt{isronlyll} parton distribution function\footnote{In the current version of MadGraph, this pdf only works with electron beams. To model a muon collider, we manually change the electron mass to that of the muon.} with a factorization scale $\mu_\text{F}=25~\text{GeV}$.

We pass the parton-level events to \texttt{Pythia8.3} \cite{Sjostrand:2014zea, Bierlich:2022pfr} to simulate \acro{QED} final-state radiation. 
We use \texttt{Delphes-v3.5.0}~\cite{deFavereau:2013fsa} for parametric, fast detector simulation and use the pre-existing \texttt{delphes-card-MuonCollider.tcl} card.
This detector card is a hybrid of the \acro{FCC}-hh and \acro{CLIC} detector recommendations~\cite{Zimmermann:2020ahu, Leogrande:2019qbe} which we take as benchmarks for the detectors at a future muon collider. 
It parameterizes the \acro{ECAL} energy resolution as
\begin{align}
    \sigma^2(E, |\eta|) &= (\sigma^{\mathrm{stochastic}}_E)^2 + (\sigma^{\mathrm{constant}}_E)^2 
    &
    \sigma^2(E, |\eta|) &= E \; a^2 + E^2 \; b^2 \label{eq:detector:resolution:parameterization}
\end{align} 
where the parameter $b = 0.01$ and the value of $a$ depends on the detector region,
\begin{align}
    a &= 0.156~\mathrm{GeV}^{\frac{1}{2}}
    &
    |\eta| &\leq 0.78  
    &&
    \text{(Barrel Region)}
    \\
    a &= 0.175~\mathrm{GeV}^{\frac{1}{2}}
    &
    0.78 < |\eta| &\leq 0.83 
    &&
    \text{(Transition Region)}
    \\
    a &= 0.151~\mathrm{GeV}^{\frac{1}{2}}
    &
    0.83 < |\eta| &\leq 2.5 
    &&
    \text{(Endcap Region)}
    \ .
\end{align}
After detector-level reconstruction, we select events according the the following criteria:
\begin{itemize}
    \item Radiative return: we select events with at least two electrons and one visible photon.
    \item \acro{VBF}: we require at least two electrons and veto events containing photons.  These include `true' \VBF{} events and radiative return diagrams with an unobservable photon.
\end{itemize}

\subsection{Kinematic Observable Distributions}
\label{sec:obs:distributions}
 
\begin{figure}[t]
    \centering
    \includegraphics[width=\linewidth]{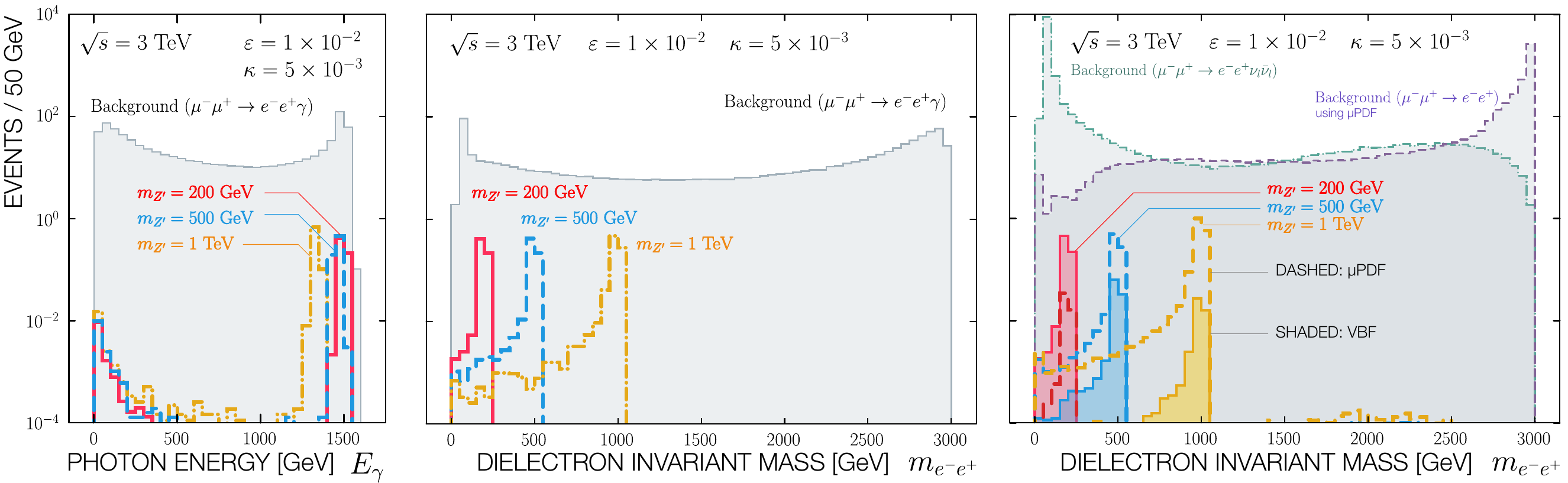}
    \caption{%
    Normalized distributions of the the photon energy $E_{\gamma}$ \textsc{(left)} and dielectron invariant mass $m_{ee}$ \textsc{(middle)} in the radiative return channel for a set of benchmark dark $Z'$ masses at a 3 TeV muon collider. 
    Normalized distributions of the dielectron invariant mass $m_{ee}$ in the \VBF{} channel (\textsc{right}) including the soft-photon limit of radiative return events ($\mu$\acro{PDF}). The shape of the latter events is described in Appendix~\ref{app:RR:contribution:to:VBF}.
    }
    \label{fig:kinematic:features:and:resolutions} 
\end{figure}

Figure~\ref{fig:kinematic:features:and:resolutions} 
presents the dielectron invariant mass and photon energy distributions for a set of benchmark $Z'$ masses in the radiative return and \VBF{} channels. The $m_{ee}$ distribution exhibits a pronounced $Z'$ resonance peak for both channels where the width increases with $m_{Z'}$. This comes from the degradation of detector resolution for energetic electrons produced in heavier $Z'$ decays shown in Figure~\ref{fig:detector:resolutions}.

\paragraph{Radiative return}
In this channel, the photon energy is another observable that may be used to discriminate signal from background. The position of the photon energy peak correlates with the dark $Z'$ mass since it encodes the energy radiation required to reach the $Z'$ resonance. Observe that in Figure~\ref{fig:detector:resolutions} the photon energy and dielectron invariant mass resolutions are complementary: at low dark $Z'$ masses the invariant mass is well resolved (small $\sigma_{m_{ee}}$) while the photon energy has poorer resolution (large $\sigma_{E_\gamma}$). The relative resolution strengths swap for heavier dark $Z'$. This observation is the basis of the optimized recoil mass technique for radiative return in muon colliders~\cite{Cheung:2025uaz}. The Standard Model background events are largely featureless in the target mass range with slight increases at low energies near the electroweak scale and near the center of mass energy where the photon is soft.

\paragraph{VBF}
For `true' \VBF{} events, the neutrinos are emitted primarily along the beamline and do not contribute large missing energy signals. The Standard Model background events with neutrinos are peaked at low dielectron invariant mass, while those without neutrinos are peaked at dielectron invariant masses close to the collider center-of-mass energy, $\sqrt{s}$. 

For heavy dark $Z'$s, the \VBF{} contribution is dominated by events from radiative return diagrams with unobservable photons, see also Figure~\ref{fig:Cross Sections}. These feature a long off-shell tail for lower dielectron invariant masses. Appendix~\ref{app:RR:contribution:to:VBF} motivates this shape. Our analysis focuses on the on-shell contribution; we leave the additional discovery reach using the off-shell tail to future work.

\subsection{Selection Cuts}
\label{sec:detector:cuts}
We tailor our analysis cuts to the topology and kinematic features of each production channel. These are as follows, subject to the pre-selection cuts in Section~\ref{sec:sig:bg:pre}: 
\begin{enumerate}
    \item Events must have at least two electrons, $N_e\geq 2$.
    \item Radiative return events must have at least one photon, $N_\gamma \geq 1$.
    \item Radiative return events must have a photon energy within the detector resolution of the expected photon energy for a dark $Z'$ of the assumed mass, $|E_{\gamma}^{Z'} - E_\gamma| < 2\sigma_{E_{\gamma}^{Z'} }$.
    \item The invariant dielectron mass must be within the detector resolution of the assumed dark $Z'$ mass, $|m_{Z'} - m_{ee}| < 2 \sigma_{m_{ee}}$.
\end{enumerate}
Our window cuts assume prior knowledge of the dark $Z'$ mass.  This follows other analyses of new heavy gauge boson resonance searches at a muon collider~\cite{Cheung:2025uaz,Cipressi:2026wtb}.
As highlighted in Ref.~\cite{San:2022uud}, dark $Z'$ bosons with the mixing parameters examined here have a narrow width and so the observed width of their resonance is set by the detector resolution. Our choices of $m_{Z'}$-dependent window cuts on the dielectron invariant mass and the photon energy optimize the signal-to-background acceptance at each hypothesis mass.

Ref.~\cite{San:2022uud} assumes that a dark $Z'$ in the mass range accessible to a future 500~GeV linear collider would have already have been observed as a bump at the \acro{LHC} that identifies the dark $Z'$ mass. This is not the case for the heavier masses accessible to a future muon collider. Our presentation of mass-dependent window cuts assumes that one performs a bump hunt scan over dark $Z'$ masses where one chooses the optimal window cut sizes for each hypothesis mass. 

\begin{figure}[t]
    \centering
    \includegraphics[width=\linewidth]{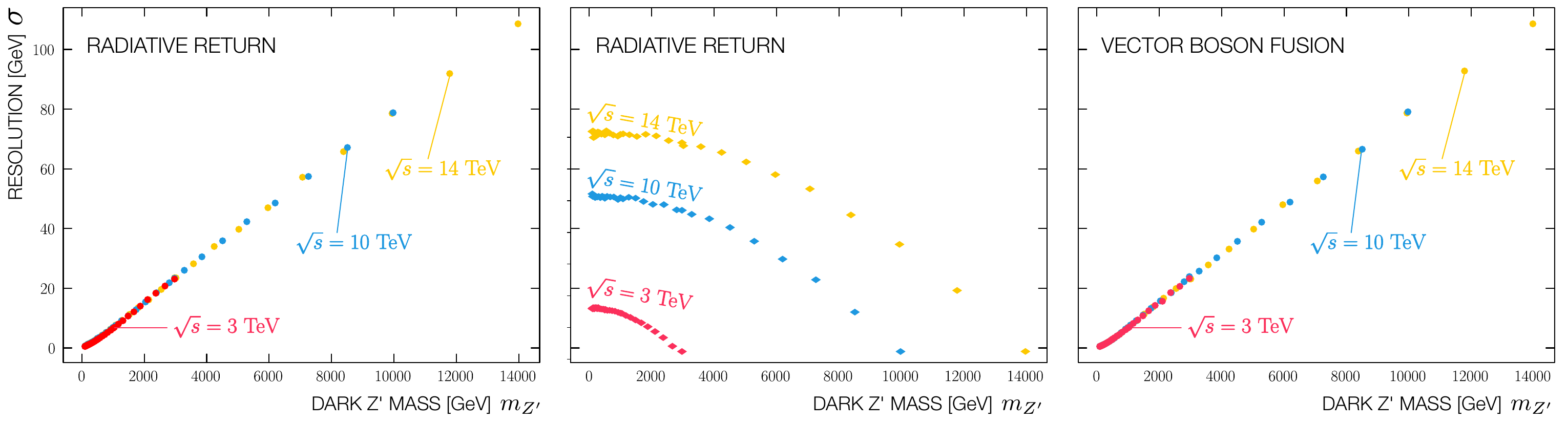}
    \caption{Resolutions for reconstructed dielectron invariant mass \textsc{(left)} and photon energy \textsc{(middle)} in the radiative return channel. Resolutions for reconstructed dielectron invariant mass \textsc{(right panel)} in the \VBF{} channel. These are shown as a function of the dark $Z'$ mass $m_{Z'}$ for benchmark center-of-mass energies of 3 TeV, 10 TeV, and 14 TeV.
    }
    \label{fig:detector:resolutions}
\end{figure}

We estimate the detector resolution for the reconstructed photon energy $E_{\gamma}$ and dielectron invariant mass $m_{ee}$ by fitting a Gaussian function to the fast detector simulation model in~\eqref{eq:detector:resolution:parameterization}.
We plot the inferred detector resolutions in Figure~\ref{fig:detector:resolutions}. These resolutions are independent of the kinetic and mass mixing parameters as the intrinsic width of $Z'$ is much smaller than the resolution of detector. As a result, the aforementioned parameters do not affect the shape of kinematic distributions.

The dielectron invariant mass resolution degrades ($\sigma_{m^{Z'}_{ee}}$ increases) as the dark $Z'$ mass increases. 
On the other hand, the photon energy resolution improves ($\sigma_{E^{Z'}_{\gamma}}$ decreases) with increasing dark $Z'$ mass since the photon energy decreases in this regime and the relative energy resolution $\sigma_{E^{Z'}_{\gamma}}/E_\gamma \sim \sqrt{E_\gamma}$~\cite[\S 35.10.2]{ParticleDataGroup:2024cfk}. As highlighted in the optimal recoil mass technique~\cite{Cheung:2025uaz}, the dielectron invariant mass and photon energy thus provide complementary information in the radiative return channel. This feeds into our counting analysis because either the dielectron invariant mass or the photon energy will more stringently reduce the relatively smooth background depending on the  dark $Z'$ mass.

Tables~\ref{tab:cut_efficiencies_1} and \ref{tab:cut_efficiencies_2} summarize the signal and background cut efficiencies for a set of benchmark dark $Z'$ models. The selection cuts suppress background events by approximately four orders of magnitude.

\begin{table}
\centering
\footnotesize
\begin{tabular}{l Lll Lll}
    \toprule
    \textbf{Selection Criteria}
    & 
    \multicolumn{3}{l}{\textbf{Background Efficiency}}
    & 
    \multicolumn{3}{l}{\textbf{Signal Efficiency}}
    \\
    $Z'$~Mass [GeV]:
    & 200 & 500 & 1000
    & 200 & 500 & 1000
    \\
    \midrule
    \addlinespace[1em]
    \multicolumn{7}{l}{$\sqrt{s} = 3$~TeV}
    \\
    \addlinespace[.5em]
    $N_e \geq 2$, $N_\gamma \geq 1$ 
    & 0.45 & 0.45 & 0.45 
    & 0.39 & 0.41 & 0.44 
    \\
    $|E^{Z'}_{\gamma} - E_{\gamma}| < 2 \sigma_{E^{Z'}_{\gamma}}$ 
    & $9.62 \times 10^{-2}$ & $5.78 \times 10^{-2}$ & $1.15 \times 10^{-2}$ 
    & 0.37 & 0.39 & 0.41 
    \\
    $|m_{Z'} - m_{ee}| < 2 \sigma_{m^{Z'}_{ee}}$ & 
    $3.50 \times 10^{-4}$ & $9.61 \times 10^{-4}$ & $1.49 \times 10^{-3}$ 
    & 0.28 & 0.29 & 0.32 
    \\
    \midrule
    \addlinespace[1em]
    \multicolumn{7}{l}{$\sqrt{s} = 10$~TeV}
    \\
    \addlinespace[.5em]
    $N_e \geq 2$, $N_\gamma \geq 1$ 
    & 0.35 & 0.35 & 0.35 
    & 0.38 & 0.37 & 0.38 
    \\
    $|E^{Z'}_{\gamma} - E_{\gamma}| < 2 \sigma_{E^{Z'}_{\gamma}}$ 
    & $3.61 \times 10^{-2}$ & $3.77 \times 10^{-2} $ & $ 4.16 \times 10^{-2}$ 
    & 0.35 & 0.35 & 0.36 
    \\
    $|m_{Z'} - m_{ee}| < 2 \sigma_{m^{Z'}_{ee}}$ 
    & $1.45 \times 10^{-4}$ & $5.64 \times 10^{-4}$ & $1.01 \times 10^{-3}$ 
    & 0.29 & 0.27 & 0.28 
    \\
    \midrule
    \addlinespace[1em]
    \multicolumn{7}{l}{$\sqrt{s} = 14$~TeV}
    \\
    \addlinespace[.5em]
    $N_e \geq 2$, $N_\gamma \geq 1$  
    & 0.46 & 0.46 & 0.46 
    & 0.37 & 0.37 & 0.37 
    \\
    $|E^{Z'}_{\gamma} - E_{\gamma}| < 2 \sigma_{E^{Z'}_{\gamma}}$ 
    & $4.39 \times 10^{-2}$ & $4.5 \times 10^{-2}$ & $4.85 \times 10^{-2}$ 
    & 0.35 & 0.35 & 0.34 
    \\
    $|m_{Z'} - m_{ee}| < 2 \sigma_{m^{Z'}_{ee}}$ 
    & $4.90 \times 10^{-5}$ & $3.37 \times 10^{-4}$ & $7.8 \times 10^{-4}$ 
    & 0.29 & 0.28 & 0.27 \\
    \bottomrule
\end{tabular}
\caption{%
    Cut flow table for the radiative return channel showing the signal and background event efficiencies after applying selection criteria.
    Results are shown for three benchmark dark $Z'$ mass values with kinetic and mass mixing parameters set to $\varepsilon = 1 \times 10^{-2}$ and $\kappa = 5 \times 10^{-3}$. 
    %
    %
    }
\label{tab:cut_efficiencies_1}
\end{table}

\begin{table}
\centering
\footnotesize
\begin{tabular}{l Lll Lll}
    \toprule
    \textbf{Selection Criteria }
    & 
    \multicolumn{3}{l}{\textbf{Background Efficiency}}
    & 
    \multicolumn{3}{l}{\textbf{Signal Efficiency}}
    \\
    $Z'$~Mass [GeV]:
    & 200 & 500 & 1000
    & 200 & 500 & 1000
    \\
    \midrule
    \addlinespace[1em]
    \multicolumn{7}{l}{$\sqrt{s} = 3$~TeV}
    \\
    \addlinespace[.5em]
    $N_e \geq 2$
    & 0.52 & 0.52 & 0.52 
    & 0.43 & 0.41 & 0.45 
    \\
    $|m_{Z'} - m_{ee}| < 2 \sigma_{m^{Z'}_{ee}}$ 
    & $2.22 \times 10^{-4}$ 
    & $1.57 \times 10^{-4}$ 
    & $1.73 \times 10^{-4}$ 
    & 0.31 & 0.31 & 0.34 \\
    \\
    \midrule
    \addlinespace[1em]
    \multicolumn{7}{l}{$\sqrt{s} = 10$~TeV}
    \\
    \addlinespace[.5em]
    $N_e \geq 2$ 
    & 0.51 & 0.51 & 0.51 
    & 0.36 & 0.30 & 0.30 
    \\
    $|m_{Z'} - m_{ee}| < 2 \sigma_{m^{Z'}_{ee}}$ 
    & 
    $2.29 \times 10^{-4}$ 
    & $1.28 \times 10^{-4}$ 
    & $9.75 \times 10^{-5}$ 
    & 0.26 & 0.22 & 0.23
    \\
    \midrule
    \addlinespace[1em]
    \multicolumn{7}{l}{$\sqrt{s} = 14$~TeV}
    \\
    \addlinespace[.5em]
    $N_e \geq 2$ 
    & 0.50 & 0.50 & 0.50 
    & 0.32 & 0.27 & 0.27 
    \\
    $|m_{Z'} - m_{ee}| < 2 \sigma_{m^{Z'}_{ee}}$ 
    & $2.16 \times 10^{-4}$ 
    & $1.47 \times 10^{-4}$ 
    & $9.30 \times 10^{-5}$ 
    & 0.24 & 0.21 & 0.21 \\
\bottomrule
\end{tabular}
    \caption{Same as Table~\ref{tab:cut_efficiencies_1} but for the VBF channel.}
    \label{tab:cut_efficiencies_2}
\end{table}

\subsection{Muon Collider Resonance Search}
\label{sec:significance}

We estimate the muon collider reach to discover a dark $Z'$ as a resonance in the radiative return and \VBF{} channels. 
We determine the signal and background cut efficiencies for each benchmark scenario and calculate the statistical significance, $\mathcal S$, for a pure counting experiment, 
\begin{align}
    \mathcal{S}^2 
    &= 
    2
    \left({N}_{\text{S}} 
        + {N}_{\text{B}}  
    \right) 
    \log
    \left(
        1 
        + \frac{{N}_{\text{S}}}{{N}_{\text{B}}}  
    \right)  
    - 
    2
    {N}_{\text{S}} 
    &
    N_i &= \sigma_i \varepsilon^i_\text{cut} \mathcal L
    \label{eq:significance}
\end{align}
where $N_\text{S}$ and $N_\text{B}$ are respectively the number of signal and background events for luminosity $\mathcal L$~\cite{Cowan:2010js}.
These, in turn, are related to their respective cross sections $\sigma_{\text{S,B}}$ and cut efficiencies $\varepsilon_\text{cut}^{\text{S,B}}$. 
To compute the significance in $\left(m_{Z'},\; \varepsilon  \right)$ plane, we obtain the signal cross sections by performing a scan over dark $Z'$ mass benchmark points and kinetic mixing parameter ranging from 
\begin{align}
    1 \times 10^{-4} &\leq \varepsilon \leq 9 \times 10^{-2}
    &
    \kappa &= 0, \; 0.5 \times 10^{-2}, \; 1 \times 10^{-2} \ .
\end{align}
We define the reach for these searches to be $\mathcal{S}=2$, corresponding to a 95\% confidence level.

\subsection{Relative Couplings and Left--Right Asymmetry}
\label{sec:relative:couplings:LR}

When resonantly producing new particle that has only one type of mixing, determining the signal cross section determines the mixing parameter. In the dark $Z'$ model, one requires further analysis to disentangle the two types of mixing ($\varepsilon, \kappa$). 

\paragraph{Complementarity of kinetic and mass mixing}
We highlight the complementary effects of the kinetic and mass mixing in Section~\eqref{sec:darkZ:EFT}. The effective coupling $g_{Z'}$ of the $Z'$ to muons \eqref{eq:g:Zp:definition} depends on a linear combination of $\varepsilon$ and $\kappa$ wherein the latter is suppressed by a ratio of the $Z$ and $Z'$ squared masses. Further, the vector-like fermion coupling carries an additional electromagnetic shift that depends only on $\varepsilon$, \eqref{eq:Z:charges}. These effects manifest themselves in the radiative return channel that produces the $Z'$ from its coupling to muons.

On the other hand, the $WWZ'$ vertex of the \VBF{} channel depends on a different linear combination of $\varepsilon$ and $\kappa$ where the mass mixing term no longer has a mass ratio penalty~\eqref{eq:delta:zeta:definition}. Furthermore, the presence of $W$ bosons `in the     ' depends on the left-chiral muons (right-chiral anti-muons) in the beam. This opens the possibility that beam polarization can help disambiguate the effects of the two types of mixing. 
Some models may have a large fraction of tagged \VBF{} events that actually depend on the $Z'$ coupling to muons---these are the `fake' \VBF{} signals in Section~\ref{sec:Zp:production:modes}. The addition of these events do not reduce the statistical reach of the number of `true' \VBF{} events in the sample. 
While it is a challenge to produce polarized muon beams, the muons injected into an accelerator are initially left-handed due to their production from weak pion decays. 
Ref.~\cite{San:2022uud} examined the role of polarized electron beams to study the dark $Z'$ model. Our treatment here follows the formalism in Appendix~D of that study, which in turn draws on the more detailed review in Ref.~\cite{Moortgat-Pick:2005jsx}. 

\paragraph{Left-Right Asymmetry}
The pure left--right asymmetry $A_\text{LR}$ is the difference of the cross sections for colliding left-handed versus right-handed initial states normalized by the total cross section. A realistic beam has a polarization fraction $P$ that is the difference in the fraction of right- and left-handed particles.\footnote{As per standard convention, a beam that is 55\% left-handed and 45\% right-handed has polarization $P= 10\%$.} For a muon beam with polarization fraction $P^{-}$ and an anti-muon beam with polarization fraction $P^{+}$, the left--right asymmetry is
\begin{align}
    A_{\text{LR}} 
    &= 
    \frac{1}{P_{\text{eff}}} 
    \frac{ 
        \sigma_{-+} - \sigma_{+-}
         }{ 
         \sigma_{-+} + \sigma_{+-}
         }
\end{align}
where the effective polarization is $P_{\text{eff}} = \left(P^{-} - P^{+} \right)\left( 1 - P^{-} P^{+} \right)^{-1}$ and the partially polarized cross sections are
\begin{align}
    \sigma_{-+} 
    &= 
    \frac{1}{4} 
    \left[
        \left(
            1 + |P^{-}| |P^{+}| 
        \right) 
        \left(
            \sigma_{\text{LR}} + \sigma_{\text{RL}} 
        \right) 
        + 
        \left(
            |P^{-}| + |P^{+}| 
        \right) 
        \left(
            \sigma_{\text{LR}} - \sigma_{\text{RL}} 
        \right)
    \right] 
    \\
    \sigma_{+-} 
    &= 
    \frac{1}{4} 
    \left[
        \left(
            1 + |P^{-}| |P^{+}| 
        \right) 
        \left(
            \sigma_{\text{LR}} + \sigma_{\text{RL}} 
        \right) 
        - 
        \left(
            |P^{-}| + |P^{+}| 
        \right) 
        \left(
            \sigma_{\text{LR}} - \sigma_{\text{RL}} 
        \right)
    \right] \ .
\end{align}
$\sigma_{\text{LR},\text{RL}}$ are the perfectly polarized cross sections. 
We assume benchmark polarization of $|P|= 30\%$ used in Refs.~\cite{Franceschini:2022sxc,Accettura:2023ked,Korshynska:2024suh}.

\section{Results}
\label{sec:Results}

We present the reach of a hypothetical high-energy muon collider to 
    (1) discover a dark $Z'$ resonance and 
    (2) determine the relative kinetic and mixing parameters.
We apply this to the choice of collider energies and luminosities in Section~\ref{sec:collider:benchmarks}.

\subsection{Discovery Reach}

\begin{figure}
    \centering
    \includegraphics[width=\linewidth]{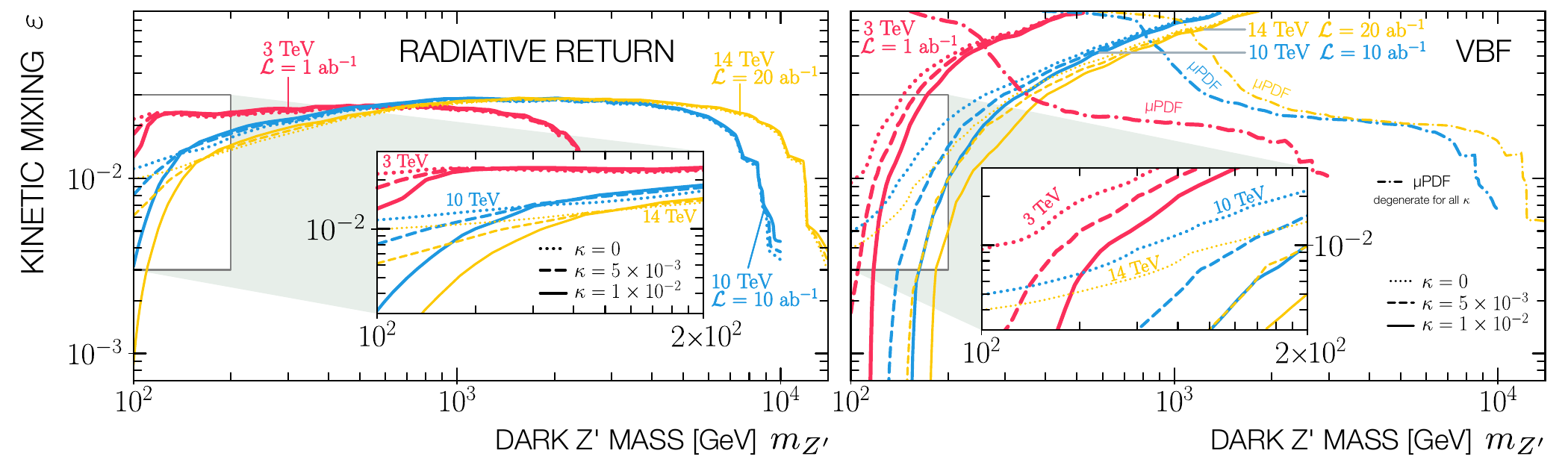}
    \caption{%
    95\% confidence level exclusion limits on kinetic mixing $\varepsilon$ as function of dark $Z'$ mass $m_{Z'}$ for the radiative return (\textsc{left}) and \VBF{} (\textsc{right}, including $\mu$\acro{PDF} contributions to \acro{VBF}) channels. We show the results for a set of muon collider collision energies (colors) and mass mixings $\kappa$ (line dashing).
    The insets zoom into the low-mass region.
    The true \VBF{} cross section for masses heavier than $m_{Z'}\gtrsim$~TeV have uncertainties larger than $\mathcal O(20\%)$ due to Sudakov double logarithms in virtual corrections, see Section~\ref{sec:on:EVA}; however in this region the \VBF{} signal is dominated by the $\mu$\acro{PDF} contribution that does not carry this theory uncertainty.
    }
    \label{fig:RRvsVBF}
\end{figure}

We plot the discovery reach as the minimum kinetic mixing  $\varepsilon$ to which one may identify a dark $Z'$ resonance over background with 95\% confidence. We plot this as a function of dark $Z'$ masses $m_{Z'}$ and for a set of benchmark mass mixing parameters $\kappa$ for ease of comparison to dark photon plots with a single type of mixing. 

Figure~\ref{fig:RRvsVBF} compares the reach in the radiative return signal of a dilepton resonance and a hard photon to the \VBF{} signal with only the dilepton resonance. As explained in Section~\ref{sec:obs:distributions}, these signatures are named to correspond to their production modes. 
The radiative return channel is relatively flat over most of the range of $m_{Z'}$ masses with increased sensitivity for light masses and masses close to $\sqrt{s}$. For light $m_{Z'}$ this is a combination of (1) background reduction due to the energetic photon requirement, (2) the stronger dielectron invariant mass resolution. For $m_{Z'}$ clear to the beam energy $\sqrt{s}$, this increased sensitivity is due to the softness of the emitted photon wherein the process effectively becomes resonant $2\to 1$ production, as we see in the production cross section in Figure~\ref{fig:Cross Sections}.

\begin{figure}
    \centering
    \includegraphics[width=\linewidth]{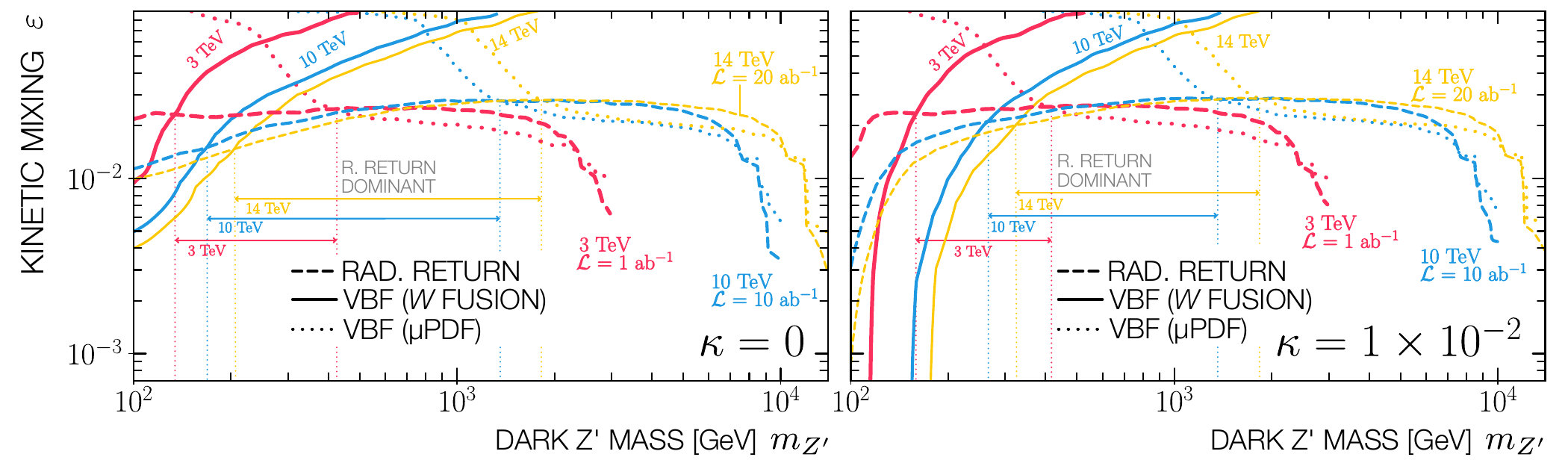}
    \caption{
    95\% confidence level exclusion limits on kinetic mixing $\varepsilon$ as function of dark $Z'$ mass $m_{Z'}$ for two illustrative values of mass mixing, $\kappa=0$ (\textsc{left}) and $\kappa=10^{-2}$ (\textsc{right}).
    We show the results for a set of muon collider collision energies (colors) and overlay the radiative return, `true' \acro{VBF}, and $\mu$\acro{PDF} channels (line dashing). The $\mu$\acro{PDF} lines are degenerate for all $\kappa$. 
    Horizontal bars masses indicate the range where radiative return is the dominant signal.
    }
    \label{fig:RRvsVBF:diffKappa}
\end{figure}

In the \VBF{} channel, the $\mu$\acro{PDF} contribution tracks the radiative return shape and is slightly stronger at heavier masses. Meanwhile, the true \VBF{} events have a stronger reach at low dark $Z'$ mass. This follows the effective vector boson approximation intuition that the $W$ content of the muon beam peaks at small momentum fraction. Thus \VBF{} is more likely when the required $W$ energy to reach the $Z'$ resonance is a small fraction of the beam energy. Increasing the mass mixing parameter $\kappa$ improves the \VBF{} reach (solid lines are to the right/below the dashed lines in Figure~\ref{fig:RRvsVBF}, right) because $\kappa$ enhances the $WWZ'$ coupling.
Figure~\ref{fig:RRvsVBF:diffKappa} plots the same information for $\kappa=0$ and $\kappa=10^{-2}$ with the radiative return and \VBF{} channels overlaid to show where the respective channels have the furthest reach. The $\mu$\acro{PDF} lines are effectively degenerate for all benchmark $\kappa$ due to the $\xi^2$ suppression in the fermion--$Z'$ coupling, \eqref{eq:g:Zp:definition}.

\subsection{Comparison to other colliders}

\begin{figure}
    \centering
    \includegraphics[width=0.7\textwidth]{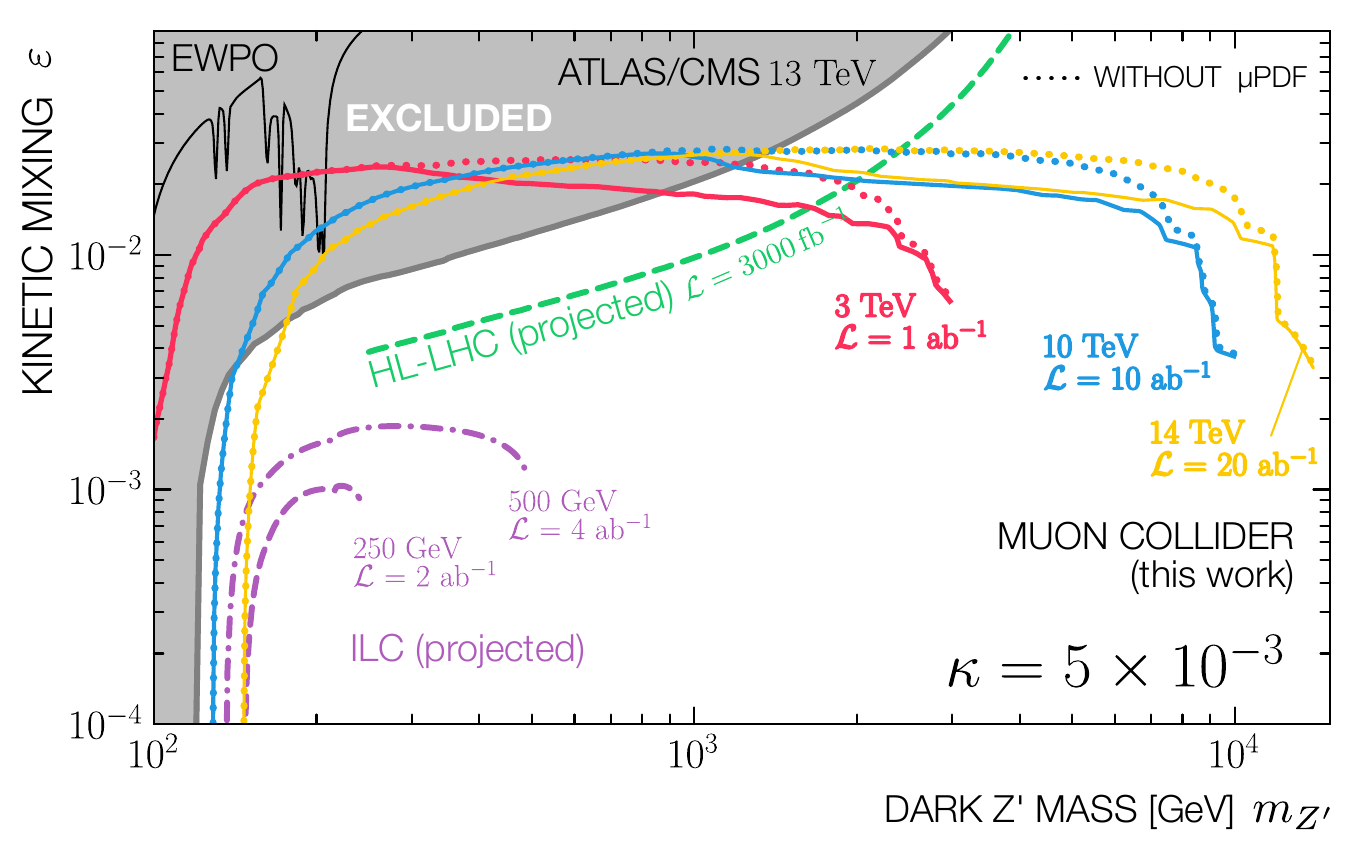}
    \caption{Combined 2$\sigma$ reach for the dark $Z'$ model for a benchmark mass mixing parameter and a set of three muon collider energies (solid, colored lines). Dotted lines show the reach if the $\mu$\acro{PDF} events are not included.
    These are compared to present exclusion (gray region) from a recast of the LHC dark photon searches (gray line)~\cite{Curtin:2014cca, Hook:2010tw, CMS:2019buh, CMS:2021ctt, ATLAS:2019erb} (via Ref.~\cite{San:2022uud}) and electroweak precision observables (black line)~\cite{Bertuzzo:2025ejw}.
    Also shown in dashed lines are the projected reach for the ILC~\cite{San:2022uud} and HL-LHC. The reach of HL-LHC is derived by Ref.~\cite{San:2022uud} from extrapolation of the existing LHC searches to the projected integrated luminosity of $\mathcal{L} = 3000 \; \text{fb}^{-1}$.
    }
    \label{fig:Sensitivities at different colliders}
\end{figure}

Figure~\ref{fig:Sensitivities at different colliders} compares the muon collider reach to that of other experiments for a benchmark value of mass mixing parameter, $\kappa = 5 \times 10^{-3}$. The \acro{LHC} creates dark $Z'$ bosons through Drell--Yan production and sets the strongest present bounds on the existence of a dark $Z'$ in $10^2$--$10^4$~GeV mass range we consider. The \acro{ILC} is an electron--positron collider that produces dark $Z'$ bosons through radiative return~\cite{San:2022uud}. However, its overall kinematic reach is limited to the available collision energy. 
A multi-TeV muon collider outperforms the \acro{LHC} for dark $Z'$ masses greater than $\mathcal O(\text{TeV})$. For heavier masses, the anti-quark \acro{PDF} is suppressed compared to a muon collider. Furthermore, we note that the \VBF{} enhancements at high energies allow a 10 TeV muon collider to outperform the \acro{LHC} in a small region above the $Z$ mass.
For the mass range we consider, electroweak precision observables are a subdominant constraint---though the power of these searches beyond the scalar decoupling limit that we assume was recently highlighted in Ref.~\cite{Bertuzzo:2025ejw}

\subsection{Determination of Mixing}

If a dark $Z'$ is discovered at a future muon collider, one must then disentangle the relative kinetic versus mass mixing. 
We use the following observables:
\begin{itemize}
 \item The unpolarized radiative return total cross section.
 \item The unpolarized \VBF{} total cross section.
 \item The left--right asymmetry, $A_{\text{LR}}$, for  polarized beams; see Section~\ref{sec:relative:couplings:LR}.
\end{itemize}
For a given luminosity $\mathcal L$, their respective uncertainties are~\cite{Korshynska:2024suh, Leike:1996pj}
\begin{align}
    \Delta \sigma_{\text{unpol}} 
    &= 
    \frac{ \sigma_{\text{unpol}} 
        }{ \sqrt{ \sigma_{\text{unpol}}  \mathcal{L} }
        } 
        &
    \Delta A_{\text{LR}} 
    &= 
    \sqrt{ 
        \frac{
            1 
            - 
            \left( 
                P_{\text{eff}} A_{\text{LR}} 
            \right)^2
        }{
                \left(
                    \sigma_{-+} + \sigma_{+-} 
                \right) 
                \mathcal{L} 
            P^2_{ \text{eff}} 
        }
    } \ .
\end{align}
We perform a $\chi^2$ analysis in the $(\kappa, \varepsilon)$-plane for a set of benchmark dark $Z'$ masses, $m_{Z'} = 200~\text{GeV}, \; 500~\text{GeV}, \; 1~\text{TeV}$.
The test statistic is~\cite[Eq.~(11) \& App.~B]{Korshynska:2024suh}
\begin{align}
    \chi^2(\kappa, \varepsilon) 
    &= 
    \sum_{j=1}^{ n_{\text{observable}} } 
    \left[
        \frac{ \mathcal{O}_{j}^{\text{model}}
                (   \kappa_{\text{b}}, 
                \varepsilon_{\text{b}}
                ) 
                - 
                \mathcal{O}_{j}(\kappa, \varepsilon)
            }{
                \Delta \mathcal{O}^{\text{model}}_{j}
                (   \kappa_{\text{b}}, 
                    \varepsilon_{\text{b}}
                )
            }
    \right]^2  
    \ ,
\end{align}
where 
    $n_{\text{observable}}=3$ is the number of observables used to construct the $\chi^2$-analysis, 
    $\mathcal{O}^{\text{model}}_{j}(\kappa_{\text{b}}, 
                \varepsilon_{\text{b}}) $ 
    is the $j^\text{th}$ observable for some benchmark values of $\kappa$ and $\varepsilon$, 
    $\mathcal{O}_{j}$ is the $j^\text{th}$ observable as a function of $\kappa$ and $\varepsilon$, 
    and $\Delta \mathcal{O}^{\text{model}}_{j}$ is the uncertainty in the $j^\text{th}$ observable.
To discriminate between dark $Z'$ models with and without mass mixing, we consider two benchmark parameter points, 
\begin{align}
    \text{\acro{BP1}}:\; (\varepsilon, \kappa)
    &= (5 \times 10^{-3}, 5 \times 10^{-3})
    &
    \text{\acro{BP2}}:\; (\varepsilon, \kappa)
    &= (5 \times 10^{-3}, 0)\ .
    \label{eq:benchmark:models}
\end{align}
The dark $Z'$ model has three parameters, the two mixing strengths and the dark $Z'$ mass. In our analysis we assume that the mass is known and fit for the two mixing parameters. We then plot the regions of fixed $\Delta\chi^2$ to determine the confidence intervals for these mixing parameters for the two benchmark points \eqref{eq:benchmark:models}. 

We do not include the Sudakov uncertainties discussed in Section~\ref{sec:on:EVA} in our projected sensitivity. This assumes that either ongoing advancements in higher-order electroweak resummation techniques for event generators or a full loop-level calculation of the matrix element will render this theory uncertainty subdominant. As such, our plots reflect the experimental reach of a future muon collider.

\paragraph{Role of Polarization}
\begin{figure}
    \centering
        \centering
        \includegraphics[width=.8\textwidth]{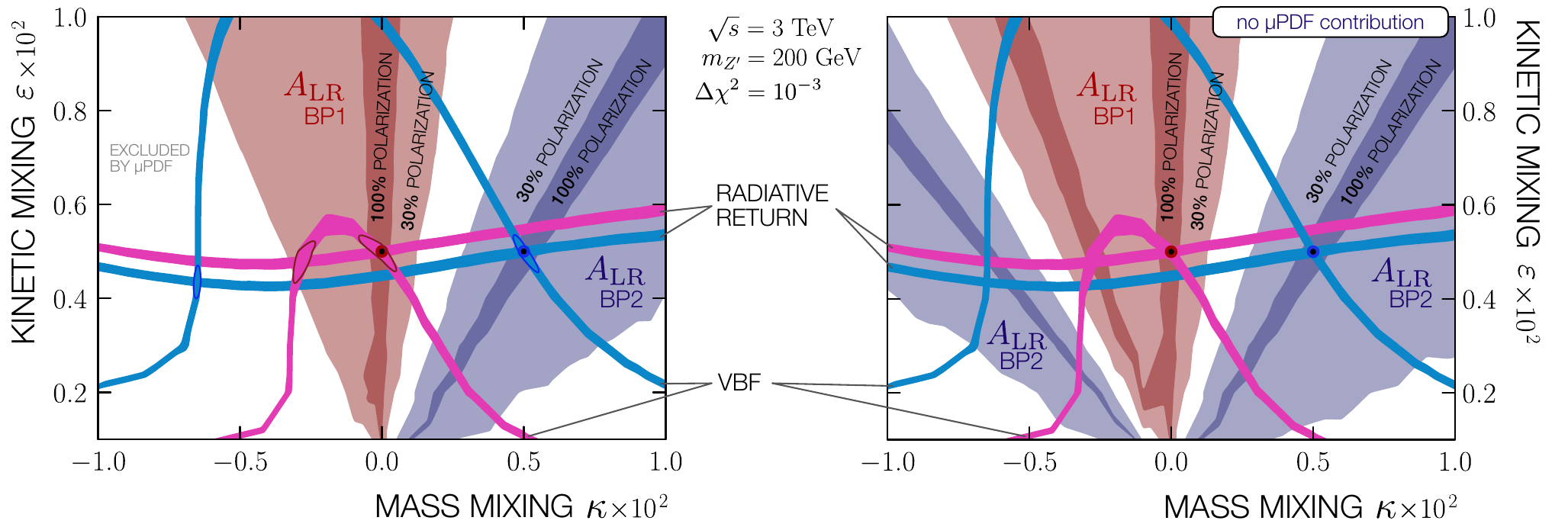}
    \caption{%
    \textsc{Left:}
    Confidence intervals for two benchmark models with respect to the unpolarized cross sections for radiative return (horizontal band), \VBF{} (wavy band), and the left--right asymmetry ($A_\text{LR}$, pizza-slices from the origin). 
    We chose an extreme value of $\Delta\chi^2 = 10^{-3}$ so that the $A_\text{LR}$ bands fit in the plot. 
    The $A_\text{LR}$ is comparable to the other observables for the unrealistic case of 100\% polarization, but is significantly weaker at 30\% polarization.
    For calibration, one may compare to the corresponding plot on the top-left of Figure~\ref{fig:chi squared tests} with $\chi^2 = 0.1$ and further the reach plot in Figure~\ref{fig:chi squared tests for different chi-squared values} which compares the combined $\chi^2=0.1$ contour to the $\chi^2 = 5.99$ ($2\sigma$) confidence interval.
    \textsc{Right:} same, but without the $\mu$\acro{PDF} events. At this minute $\Delta\chi^2$ value, the additional $\mu$\acro{PDF} events can break the degeneracy in best-fit solutions---however, for realistic values of $\Delta\chi^2$ the bands effectively cover the whole space.
    }
    \label{fig:RoleOfPolarization}
\end{figure}
We find that the left--right asymmetry from 30\% polarized beams does not significantly contribute to parameter estimation. To illustrate this, we show contours for an extreme value $\Delta\chi^2 = 10^{-3}$ in Figure~\ref{fig:RoleOfPolarization}. This value is chosen to fit the $A_\text{LR}$ confidence interval within the plot. We also plot an unrealistic ideal case of 100\% polarization, which is comparable to the unpolarized radiative return and \VBF{} cross sections. 
The lesson here is that the kinetic and mass mixing parameters are already distinguishable from their complementary effects on the radiative return and vector boson fusion unpolarized cross sections. By measuring these two cross sections---dielectron resonances with and without a hard photon---at a high-energy muon collider, one may already distinguish between $\varepsilon$ and $\kappa$. With the extreme value of $\Delta\chi^2$, the $\mu$\acro{PDF} can break the  $\mathbbm{Z}_2$ degeneracy in the best-fit parameters; this is due to the $\xi^2$ relative weighting in the muon--$Z'$ coupling \eqref{eq:g:Zp:definition}. For realistic $\Delta\chi^2$ values and our benchmark luminosities, however, the bands are too large to be statistically significant.

\paragraph{Unpolarized cross sections}
\begin{figure}[t]
    \centering
    \includegraphics[width=\linewidth]{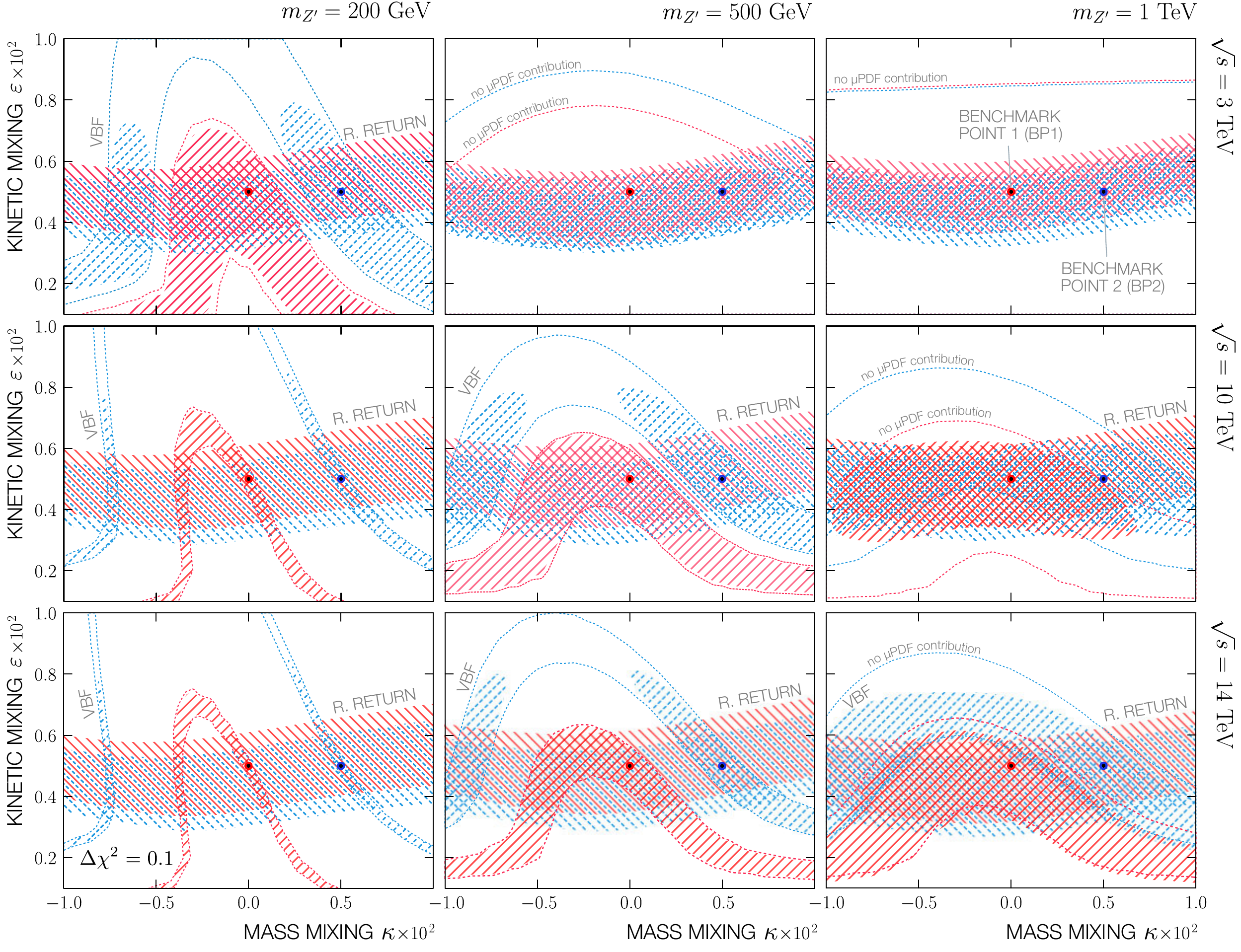} 
    \caption{%
    $\Delta \chi^2 = 0.1$ contours for the two benchmark points in \eqref{eq:benchmark:models} (\acro{BP1} in blue dashed, \acro{BP2} in red) with respect to the radiative return (north-east hash) and \VBF{} (south-west hash) channels.
    The grid of plots corresponds to three choices of collider energy (rows) and dark $Z'$ masses (columns).
    The virtual Sudakov uncertainty on the \VBF{} band width is under 10\% for each mass.
    Dotted lines show the \VBF{} band without $\mu$\acro{PDF} data.
    }
    \label{fig:chi squared tests}
\end{figure}
Figure~\ref{fig:chi squared tests} shows the $\Delta\chi^2 = 0.1$ contours for the radiative return and \VBF{} unpolarized cross sections for a set of collider energies and dark $Z'$ masses. This small value of $\Delta\chi^2$ is chosen for aid the visual comparison between different plots and is not representative of a discovery reach. 

The figure confirms that the two channels constrain different combinations of the two mixing parameters. The radiative return channel is a nearly horizontal band since the muon coupling \eqref{eq:g:Zp:definition} is dominated by the kinetic mixing. On the other hand, the \VBF{} channel depends on a coupling \eqref{eq:delta:zeta:definition} with comparable contributions from both $\varepsilon$ and $\kappa$. We highlight the role of the relative sign of the couplings by plotting negative values of $\kappa$. The asymmetry in the plots about $\kappa=0$ shows how kinetic and mass mixing may interfere either constructively or destructively.
The dashed lines show the $\Delta\chi^2 =0.1$ contours if the $\mu$\acro{PDF} events were excluded from the \VBF{} dataset. For Benchmark Point 2 with nonzero $\kappa$ we see that the effect of these additional $\mu$\acro{PDF} events is to tighten the \VBF{} contours toward the radiative return contour. This reflects the fact that these additional events depend on the same muon--$Z'$ coupling as the radiative return events. In the case of large $m_{Z'}$ and small $\sqrt{s}$ (top right) where there are many $\mu$\acro{PDF} events, the contours merge into the same band.

\paragraph{Combined Fit}
\begin{figure}[t]
    \centering
    \includegraphics[width=\linewidth]{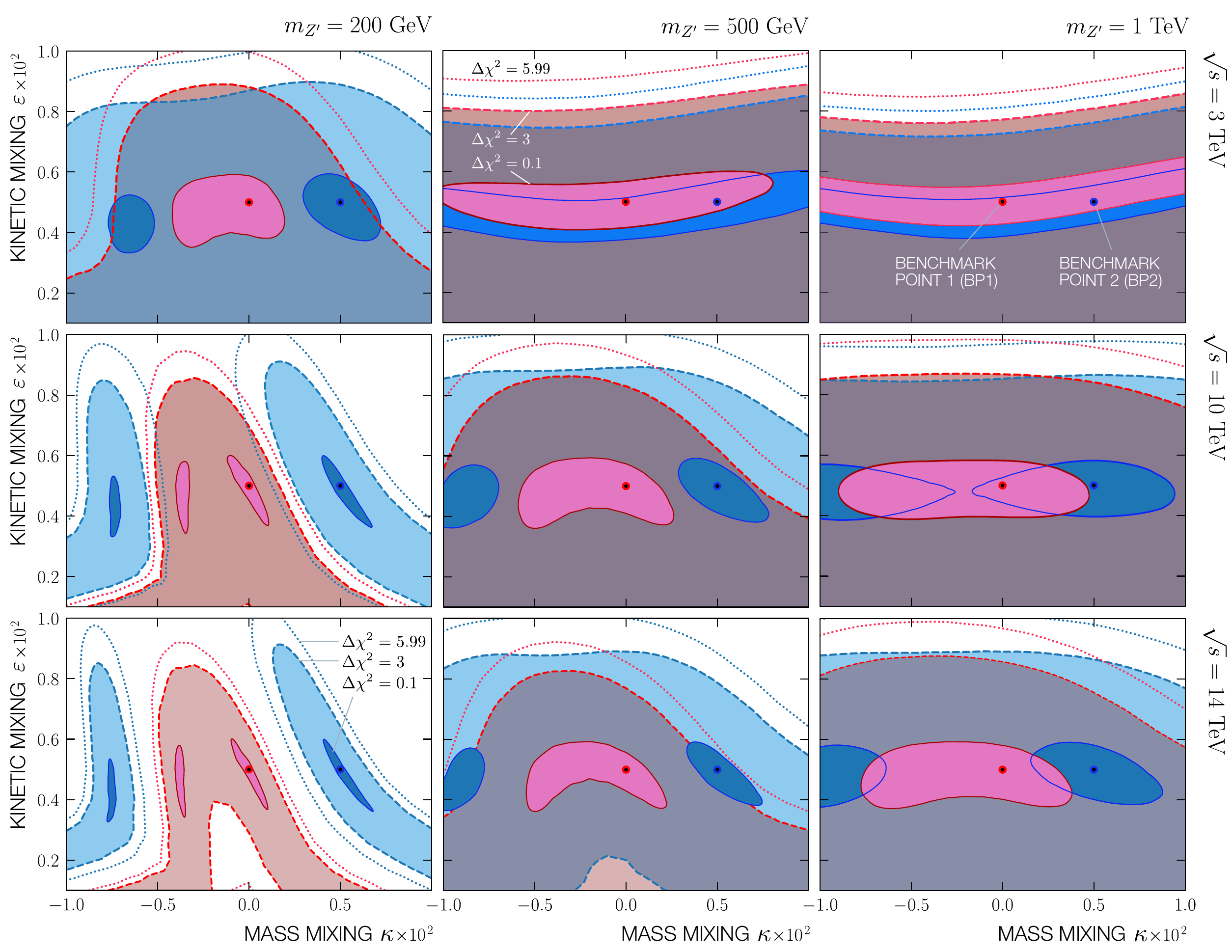}
    \caption{$\Delta \chi^2 = 0.1$ (solid), $3$ (dashed), $5.99$ (dotted) contours for different dark $Z'$ masses and collider energies for two benchmark points in \eqref{eq:benchmark:models} (\acro{BP1} in blue, \acro{BP2} in red). The outermost contour corresponds to a $2\sigma$ confidence. $A_\text{LR}$ data is included but the effect is negligible. The grid of plots corresponds to three choices of collider energy (rows) and dark $Z'$ masses (columns). 
    }
    \label{fig:chi squared tests for different chi-squared values}
\end{figure}
Figure~\ref{fig:chi squared tests for different chi-squared values} shows the confidence intervals for the combined 30\% polarized and unpolarized data. The fit is dominated by the unpolarized data; the left-right asymmetry does not make a visible impact. The outermost contours correspond to a $2\sigma$ confidence, while the innermost contours match the $\Delta\chi^2 = 0.1$ lines in Figure~\ref{fig:chi squared tests}. The appearance of degenerate islands where two values of $\kappa$ minimize $\chi^2$ illustrates the aforementioned constructive versus destructive interference between the two mixing types.

A higher-energy muon collider tightens the \VBF{} contour but does not appreciably affect the radiative return channel. This is because at higher energies the momentum fraction required to produce a $W$ boson of a given energy is smaller, and therefore it is more likely to emitting a $W$ (the $W$ distribution function) of fixed energy to reach the dark $Z'$ resonance. As the dark $Z'$ mass increases for fixed beam energy, the contours become larger since there are fewer events due to a reduced $W$ distribution function.
On the other hand, the radiative return process produces dark $Z'$ on shell by radiating a hard photon. This is mostly independent of the collider energy since that only adjusts the energy of the emitted photon.

\section{Summary and Conclusions}
\label{sec:Conclusion}

The dark $Z'$ model introduces a sequestered Abelian gauge boson that interacts with the Standard Model through kinetic and mass mixing. 
We examine the reach of a future high-energy muon collider to discover a dark $Z'$ boson and determine the relative mixing parameters. We focus on a range of masses from 100~GeV to the collider energy, which we take to be up to 14~TeV. For each collider energy, we assume a luminosity commensurate the sensitivity needs of the electroweak physics program.
 
The \acro{LHC} or a future sub-TeV lepton collider would produce such a state through its fermionic coupling---either through Drell--Yan or $\ell^+\ell^-$ annihilation in association with a hard photon (radiative return). Disentangling the relative mixing then requires tests of the $Z'$ chiral couplings, for example through left--right asymmetries if one can polarize the beam.
While radiative return is also viable channel at a muon collider, a new feature is that at high enough energies, the muon collider is effectively an electroweak vector boson collider. This accesses an alternative production mechanism, vector boson fusion $W^+W^-\to Z'$, which probes a complementary combination of the kinetic and mass mixing parameters. Unlike radiative return, \VBF{} produces a dilepton final state without an associated photon. 

We calculate the relative radiative return and \VBF{} production rates and show that \VBF{} dominates for large collider energies and relatively low dark $Z'$ mass where the $W$ momentum fraction of the beam is small. This follows intuition from effective vector boson approximation wherein the $W$ may be treated as a component of a high-energy muon beam. The range energies and masses we consider span the boundary of when this approximation is valid; we thus choose to instead simulate the full tree-level matrix element including the collinear neutrinos from $\mu\to W\nu$ splitting. We include radiative return events with a soft/collinear photon in the \VBF{} sample and show when this population drives the discovery reach.

A $\mathcal O(10~\text{TeV})$ muon collider has clear advantages in energy scale and minimal underlying event activity when comparing to hadron colliders or a lower-energy linear collider. We summarize its relative discovery reach for an example model in Figure~\ref{fig:Sensitivities at different colliders}. More surprising is that the unpolarized radiative return and \VBF{} cross sections alone are as effective in disambiguating the two types of mixing as having 100\% polarized beams; we illustrate this in Figure~\ref{fig:RoleOfPolarization}. We demonstrate the ability to separately determine kinetic and mass mixing in Figure~\ref{fig:chi squared tests for different chi-squared values} where the slight asymmetry between positive and negative values of the mass mixing parameter reflect the constructive versus destructive interference between the two mixing types.
These observations further the case for the unique reach of a future high-energy muon collider for physics beyond the Standard Model. 

There are natural future directions that merit follow up study. Some of these include:
\begin{itemize}
    \item Our study focused on the dielectron final state as a clean signature. However, for heavy dark $Z'$ and large mass mixing, the fermionic final states are subdominant to $WW$ and $Zh$ production. This motivates a more careful study of the richer set of final states and the extent to which a careful measurement of the $Z'$ branching ratios can further disambiguate the mass and kinetic mixing parameters. 

    \item The dark $Z'$ model may be a useful beyond the Standard Model benchmark for the rich and ongoing program developing the effective vector boson approximation. It realizes as simple, few-parameter model whose signatures depend on the $W$ content of \emph{each} beam.

    \item Continuing the theme of optimistic futures for high-energy physics, another possible future is a 100~TeV proton--proton collider~\cite{Arkani-Hamed:2015vfh}. Unlike the muon collider where one may distinguish radiative return from \VBF{} through the final state particles, a high-energy proton collider is expected to produce dark $Z'$s through Drell--Yan and \acro{VBF}. Neither of those channels produces an additional hard particle to tag on. It would be curious to determine the discovery reach of such a collider and to determine whether one can distinguish between the two types of mixing. 

    \item In some regimes, the \VBF{} sample is dominated by radiative return events with an unobservable photon. These have a low-invariant mass tail that may lend itself to further shape analysis beyond our cut-and-count approach here, though such an analysis will depend on yet-undetermined details of a future collider.

\end{itemize}

\section*{Acknowledgments}
We appreciate correspondence and conversations with 
    Fernanda Hüller (who provided a copy of the code for electroweak precision observables in Ref.~\cite{Bertuzzo:2025ejw}),
    Simon Knappen,
    Tania Robens,
    Richard Ruiz (for generous correspondence on muon collider physics, the effective vector boson approximation, large logarithms, and comments on a draft of this manuscript),
    Olivier Mattelaer (for tireless support answering MadGraph questions),
    Maxim Perelstein (who pointed out the significance of radiative return events with an unobservable photon),
    Yik Chuen (Eric) San,
    and
    Francis Lance Jumawan.
\acro{PT} is supported by a \acro{NSF CAREER} award (\#2045333). Portions of this work were completed at the Aspen Center for Physics (\acro{NSF} grant \acro{PHY-2210452}) and the Kavli Institute for Theoretical Physics (\acro{NSF} grant \acro{PHY-2309135}).
\acro{NAR} acknowledges the support from Department of Science and Technology - Science Education Institute (\acro{DOST-SEI}) through Accelerated Science and Technology Human Resource Development Program (\acro{ASTHRDP}) scholarship.


\appendix

\section{Model Summary}
\label{app:model:summary}

We briefly summarize the dark $Z'$ model in Ref.~\cite{San:2022uud}, which we briefly summarize here for completeness. The model posits an additional U(1)$_\text{d}$ dark gauge symmetry that mixes with the Standard Model in two ways:
\begin{enumerate}
    \item \textbf{Kinetic mixing} with hypercharge analogous to the dark photon.
    \item \textbf{Mass mixing} through the vacuum expectation value (vev) of a scalar that carries both the charges of the Standard Model Higgs and the additional Abelian symmetry. 
\end{enumerate}
We assume that additional \acro{UV} interactions decouple all additional scalar degrees of freedom other than the Goldstone bosons of the electroweak bosons and the dark $Z'$. Appendix~B of Ref.~\cite{San:2022uud} presents a minimal realization of how these states may be decoupled.

We parameterize mass mixing with an order parameter, $v_\text{mix}^2$, that simultaneously breaks electroweak and the U(1)$_\text{d}$ symmetry. A simple realization is that $v_\text{mix}^2$ is the vacuum expectation value (vev) of an additional Higgs doublet that also carries dark charge. The parameter $\kappa$ is proportional to the ratio of this order parameter to that of the Standard Model Higgs,\footnote{This ratio is known as $\tan\beta$ in the two-Higgs doublet model literature. Our Higgs sector is that of a Type~I two-Higgs doublet model at small $\tan\beta$.} $v_\text{mix}^2/v_\text{EM}^2$. We require the total order parameter of electroweak symmetry breaking to match the measured value,
\begin{align}
    v^2 &= v^2_{\text{EW}} + v^2_{\text{mix}} = (246~\text{GeV})^2
    &
    v_{\text{mix}}^2 &\ll v_{\text{EW}}^2 \ .
    \label{eq:vevs:246}
\end{align}
The mixing induces low-energy corrections to the kinetic and gauge mass terms,
\begin{align}
    \Delta \mathcal L_\text{kin}
    &= 
    \frac{\varepsilon}{2\cos\theta_\text{W}}
    B_{\mu\nu}X^{\mu\nu}
    &
    \Delta \mathcal L_\text{mass}
    &=
    \left(
        -
        \frac{1}{2}
        g_Z Z
        + g_\text{D} X 
    \right)^2 
    \frac{v_\text{mix}^2}{2}
    \ ,
\end{align}
where $X_\mu$ and $X_{\mu\nu}$ are the new gauge field and its field strength and $g_Z Z = -g' B + g W^3$. We absorb the mixed Higgs dark charge into $v_\text{mix}$.

\subsection{Diagonalizing the Lagrangian}

To find the gauge boson mass eigenstates, one first diagonalizes the kinetic mixing,\footnote{The kinetic mixing is diagonalized by an \acro{SO(2)} rotation between the hypercharge and dark gauge boson. The kinetic terms, however, are no longer canonically normalized so that one must rescale the gauge fields. After this, the kinetic terms are not only diagonal, but they are universal---invariant under any additional \acro{SO(2)} rotations. When the dark gauge field has a mass term but no mass mixing, the obvious basis choice is one with a massless hypercharge boson and a massive dark gauge boson. This gives the transformation \eqref{eq:app:kin:mix}, wherein any additional dark sector particles do not pick up any Standard Model charges while Standard Model currents pick up a $\varepsilon$-suppressed dark charge.}
\begin{align}
    B_\text{old} 
    &= 
    B_\text{new} 
    + 
    \frac{\varepsilon}{\cos\theta_\text{W}} X_\text{new}
    \ .
    \label{eq:app:kin:mix}
\end{align}
We work only to leading order in the mass and kinetic mixing parameters. The mass mixing term can then be diagonalized with respect to the Standard Model $Z$-mass term $\frac{1}{4}g_Z^2 v_\text{EW}^2 Z^2$ and a pure dark sector mass term, $\frac{1}{2} {M_X^2}X^2$. The total order parameter for electroweak symmetry breaking must be the same as the Standard Model, \eqref{eq:vevs:246}.
%
It is convenient to parameterize the mass mixing in a dimensionless parameter $\kappa$, and to identify a linear combination of $\kappa$ and $\varepsilon$ that appears in mass formulae:
\begin{align}
    \kappa &\defeq 2 
    \frac{ g_\text{d} }{ g_Z } 
    \frac{ v_\text{mix}^2 }{ v_\text{EW}^2 } 
    &
    \zeta &\defeq \kappa + \varepsilon\tan\theta_\text{W}
    \ .
\end{align}

The Standard Model $Z$ and the dark $Z'$ are the eigenstates of the Lagrangian mass terms,
\begin{align}
    \Delta \mathcal L_\text{mass}
    &= 
        \frac{1}{2} M_Z^2 Z^2 + 
        M_Z^2 
        \zeta\,
        Z\cdot X + 
        \frac{1}{2} M_{X}^2 X^2
    &
    m_Z^2 
    &
    \defeq
    \frac{1}{4} 
    \frac{g^2}{\cos\theta_\text{W}^2}
    v^2 
    \label{eq:model:mass:terms:pre:diagonal}
        \ ,
\end{align}
where $M_X^2$ depends on $m_X^2$, $\kappa$, and $\varepsilon$.
Denote the mass eigenvalues by $m_Z^2$ and $m_{Z'}^2$. The Standard Model $Z$ mass is unchanged from its Standard Model value up to quadratic order in $\zeta\ll 1$, $m_Z^2 = M_Z^2 + \mathcal O(\zeta^2)$. The dark $Z'$ mass, $m_{Z'}^2$, picks up a linear correction relative to $M_X^2$, but because the dark sector mass $m_X^2$ is arbitrary, we instead simply choose the physical mass $m_{Z'}^2$ to be our free parameter encoding all U(1)$_\text{d}$ symmetry breaking that is not already included in $v_\text{mix}^2$.

\subsection{UV Model and Effective Parameters}
\label{sec:UVmode:EffParams}
We refer to Ref.~\cite[Sec.~2 and App.~B]{San:2022uud} for details of an explicit ultraviolet model that realizes our scenario. The key features are as follows:
\begin{enumerate}
    \item The symmetry-breaking order parameters $v_\text{EW}$ and $v_\text{mix}$ are the vevs of a pair of Higgs doublets. $H_\text{EW}$ is effectively the Standard Model Higgs doublet while $H_\text{mix}$ is also charged under the dark sector. For simplicity, we may assume a Type-I Two-Higgs Doublet Model where only $H_\text{EW}$ couples to the Standard Model fermions and the limit of large $\tan\beta = v_\text{EW}^2/v_\text{mix}^2 \gg 1$. 

    \item We further assume a contribution to the dark $Z'$ mass term that may come from a dark sector Higgs, $H_\text{d}$, with no Standard Model quantum numbers. This makes the physical dark $Z'$ mass, $m_{Z'}$, a free parameter.

    \item A trilinear term proportional to $\mu' H_\text{EW}^\dag H_\text{mix} H_\text{d}$ explicitly breaks  global symmetries of the scalar sector lift the masses of would-be Goldstone bosons to the scale $\mu'$ that are left un-eaten by the massive gauge bosons. The resulting potential is stable in all field directions.
\end{enumerate}

We assume that the $Z'$ is the lightest dark sector degree of freedom and that all other states are decoupled. At low energies, $E\sim m_{Z'}$, the dark sector is effectively parameterized by the dark $Z'$ mass $m_{Z'}$, the kinetic mixing $\varepsilon$, and the mass mixing parameter $\kappa$.

Ref.~\cite{Bertuzzo:2025ejw} examines electroweak precision bounds on this model, which they call a dark photon with generalized mixing. In addition to the $Z'$ mass $m_{Z'}$, they use an alternative set of model parameters 
\begin{align}
    \xi_\text{Bertuzzo} &\approx \varepsilon
    &
    \tan\beta &= \frac{ v_\text{EW}^2 }{ v_\text{mix}^2 }
    = \frac{ 2g_\text{d}}{ g_Z } \frac{1}{\kappa}
    &
    \tan\eta &= \frac{ v_\text{d}^2 }{ v_\text{mix}^2 }
    \label{eq:Bertuzzo:comparison}
\end{align}
instead of the kinetic mixing $\varepsilon$, mass mixing $\kappa$. In our work, we have one fewer parameter because we fix the value of the net order parameter of electroweak symmetry breaking, \eqref{eq:vevs:246}, which is inferred from the $W$ mass and Fermi constant $G_\text{F}$ to better than one part per million~\cite{ParticleDataGroup:2024cfk}.

\section{\texorpdfstring{%
	Dark $Z$ Partial Decay Widths%
	}{Dark Z Partial Decay Widths}}
\label{appendixA}

We present analytical expressions for the partial decay widths of the dark $Z'$ to specific Standard Model particles. We work to leading order in the small parameters $\varepsilon$ and $\kappa$. We also ignore the masses of leptons, $\ell = e, \mu, \tau$, as well as up, down, charm, and strange quarks.
We present the vector-like and axial couplings of the dark $Z'$ to Standard Model fermions in \eqref{eq:gZp:V:A} with respect to the effective dark $Z'$ coupling $g_{Z'}$ defined in \eqref{eq:g:Zp:definition} and the Standard Model $Z$ charges in \eqref{eq:Z:charges}.

The decay width of $Z'$ to a massless fermion then takes the following form:
\begin{align}
	\Gamma_{Z'\to f\bar f} &= 
	\Gamma_0 
	\left[
		(g_{Z'f}^\text{V})^2 
		+ 
		(g_{Z'f}^\text{A})^2
	\right] 
	&
	\Gamma_0 &= \frac{m_{Z'}}{12\pi} \ .
\end{align}
The decay widths are as follows.
\begin{align}
	\Gamma_{ Z' \rightarrow \ell \bar{\ell} } 
	&=
	\Gamma_0
	\left[
		\left(
			-\varepsilon e
			- g_{Z'} 
			\left(-\frac{1}{4} + \text{s}_\text{W}^2\right)
		\right)^2
		+
		\left(
			-\frac{g_{Z'}}{4}
		\right)^2
	\right]
	\\
	\Gamma_{ Z' \rightarrow \nu \bar{\nu} } 
	&=
	2 \Gamma_0 \left(
			\frac{g_{Z'}}{4}
		\right)^2
	\\
	\Gamma_{ Z' \rightarrow u \bar{u}, c \bar{c}} 
	&=
	3
	\Gamma_0
	\left[
		\left(
			\frac{2}{3} \varepsilon e
			- g_{Z'} 
			\left(\frac{1}{4} - \frac{2}{3} \text{s}_\text{W}^2\right)
		\right)^2
		+
		\left(
			\frac{g_{Z'}}{4}
		\right)^2
	\right]
	\\
	\Gamma_{ Z' \rightarrow d \bar{d}, s \bar{s}} 
	&=
	3
	\Gamma_0
	\left[
		\left(
			-\frac{1}{3} \varepsilon e
			- g_{Z'} 
			\left(-\frac{1}{4} + \frac{1}{3} \text{s}_\text{W}^2\right)
		\right)^2
		+
		\left(
			-\frac{g_{Z'}}{4}
		\right)^2
	\right] 
	\ .
\end{align}
The factor of 3 on the quark final states accounts for three colors. The decay width to heavy quarks are rescalings of the associated massless quark limit to account for kinematic factors,
\begin{align}
	\Gamma_{ Z' \rightarrow t \bar{t}} 
	&= \frac{ m_{Z'}^2 + 2m_t^2 }{ m_{Z'}^2  } 
		\sqrt{1 - \frac{4m_t^2}{m_{Z'}^2} }
		\Gamma_{ Z' \rightarrow u \bar{u}} 
	\\
	\Gamma_{ Z' \rightarrow b \bar{b}} 
	&= \frac{ m_{Z'}^2 + 2m_b^2 }{ m_{Z'}^2  } 
		\sqrt{1 - \frac{4m_t^2}{m_{Z'}^2} }
		\Gamma_{ Z' \rightarrow d \bar{d}}  \ .
\end{align}
The decay widths into bosons are
\begin{align}
	\Gamma_{Z' \rightarrow Z h} 
	&= 
	\frac{ g^2_{Z'Zh} }{ 192 \pi m_{Z'} } 
	\left( 
		8 
		+  
		\frac{ m^2_{Z'} + m^2_Z - m^2_h }{ m_{Z'} m_Z } 
	\right)
	\left[ 1 
		- \left( 
			\frac{ m_Z + m_h }{ m_{Z'} }
			\right)^2 
	\right]^\frac{1}{2} 
	\left[ 1 
		- \left( \frac{m_h - m_Z}{m_{Z'} } \right)^2 
	\right]^\frac{1}{2} 
	\\
	\Gamma_{Z' \rightarrow W W} 
	&= 
	\frac{ m_{Z'} g_{Z' W W}^2   }{ 48 \pi } 
		\sqrt{ 1 - \frac{ 4 m^2_W }{ m^2_{Z'} } } 
		\left( \frac{m_{Z'}}{m_W}  \right)^4 
		\left[ 
			\frac{1}{4} 
			+ 4 \left( \frac{m_W}{m_{Z'}}  \right)^2 
			- 17 
			- 12  \left( \frac{m_W}{m_{Z'}}  \right)^6
		\right]
		\ .
\end{align}

\section{Distribution Muon PDF Events}
\label{app:RR:contribution:to:VBF}

Diagrams with the radiative return topology but whose photons are too soft and/or collinear to be detectable---for example, not passing the photon selection cuts---are not counted in the radiative return dielectron plus photon signal. Instead, these events contribute to the dielectron, no photon `\acro{VBF}' signal. They are effectively modeled by treating the muon beam has having a muon parton distribution function.

Figure~\ref{fig:kinematic:features:and:resolutions} shows the distribution of dielectron invariant masses from this population (labeled $\mu$\acro{PDF}). Unlike the true-VBF events, which are closer to a Breit--Wigner shape, these $\mu$\acro{PDF} events have a long low-invariant mass tail. These reflect off-shell $Z'$ production enhanced by the parton distribution functions. One might be curious why such a tail exists. After all, the muon \acro{PDF} is peaked at energy fraction $x\to 1$, \emph{not} at $x\to 0$. And if these are off-shell, why is there not a symmetric tail for invariant masses larger than the $Z'$ mass? This appendix addresses this shape. While the material here is well known among collider physicists, see e.g.~\cite[\S2.1]{AlAli:2021let}, we provide here for completeness. 

For a collision between partons of energy fractions $x_1$ and $x_2$, the partonic center of mass squared energy is $\hat s = x_1 x_2 s$. This is the energy of the observed dielectron invariant mass, $m_{ee}^2$. The differential cross section is an integral over these energy fractions,
\begin{align}
    \frac{\D{\sigma}}{\D{\hat s}}
    &\propto
    \int_{0}^1
    \D{x_1}\D{x_2}\, 
    f_\mu(x_1)f_{\bar\mu}(x_2)
    \frac{ 
        \delta( x_1 x_2 s - \hat{s} ) 
        }{
        (\hat{s}-m_{Z'}^2)^2 + \Gamma_{Z'}^2 m_{Z'}^2
    } 
    =
    \int_{\hat{s}/s}^1 
    \frac{ \D{x_1} }{ x_1 s }
    %
    \frac{ 
            f_\mu(x_1)
            f_{\bar\mu}(\hat{s}/x_1s)
        }{
        (\hat{s}-m_{Z'}^2)^2 + \Gamma_{Z'}^2 m_{Z'}^2
    } 
    \ .
    \label{eq:PDFs:one:over:x1}
\end{align}
On the right-hand side the lower limit of the $\D{x_1}$ integration is set by the upper limit $x_2 \leq 1$. We see that the Jacobian factor introduces an enhancement for small momentum fraction $x_1$.
Combined with the long tail in the muon \acro{PDF} $f_\mu(x)$ \cite{Garosi:2023bvq}, this tells us that off-shell $Z'$ production---events that contribute $m_{ee}^2\neq m_{Z'}^2$---prefers one muon having low energy fraction. The form of the $\D{x_1}$ integrand also shows why the off-shell tail only extends to small invariant mass: large values of $\hat{s}$ have a smaller $\D{x_1}$ integration volume and miss the largest part of the $x_1^{-1}$ Jacobian enhancement.

\bibliographystyle{utcaps} 	
\bibliography{MuonDarkZ}

\end{document}